\documentclass{aa}  

\usepackage{graphicx}
\usepackage{sidecap}
\usepackage[none]{hyphenat}
\usepackage[pdfpagelabels=false]{hyperref}
\hypersetup{colorlinks=true,linkcolor=blue,citecolor=blue,filecolor=blue,urlcolor=blue,}
\makeatletter
\renewcommand*\aa@pageof{, page \thepage{} of \pageref*{LastPage}}
\makeatother

\usepackage{adjustbox}  %
\usepackage{amsmath}	% Advanced maths commands
\usepackage{amssymb}	% Extra maths symbols
\usepackage{multirow}   % Extra command \multirow
\usepackage{array}       
\usepackage{mathtools}
\usepackage{comment}  %
\usepackage{array, makecell} %
\usepackage{longtable}
\usepackage{caption}
\usepackage{subcaption} % per mettere figure vicine con unica didascalia/legenda
\usepackage{xcolor}
\usepackage{isotope}
\definecolor{green_comm}{RGB}{0,160,0}
\definecolor{orange}{RGB}{255,165,0}

\def\TiI{\hbox{\rm Ti\,$\scriptstyle\rm I$}}
\def\VI{\hbox{\rm V\,$\scriptstyle\rm I$}}
\def\CrI{\hbox{\rm Cr\,$\scriptstyle\rm I$}}
\def\FeI{\hbox{\rm Fe\,$\scriptstyle\rm I$}}
\def\NaI{\hbox{\rm Na\,$\scriptstyle\rm I$}}
\def\MgI{\hbox{\rm Mg\,$\scriptstyle\rm I$}}
\def\LiI{\hbox{\rm Li\,$\scriptstyle\rm I$}}
\def\HeI{\hbox{\rm He\,$\scriptstyle\rm I$}}
\def\ScI{\hbox{\rm Sc\,$\scriptstyle\rm I$}}
\def\MnI{\hbox{\rm Mn\,$\scriptstyle\rm I$}}
\def\CoI{\hbox{\rm Co\,$\scriptstyle\rm I$}}
\def\CaII{\hbox{\rm Ca\,$\scriptstyle\rm II$}}

\def\kms{$\mathrm{km\,s}^{-1}$}
\newcolumntype{P}[1]{>{\centering\arraybackslash}p{#1}}
\begin{document}

   \title{High-resolution transmission spectroscopy of the hot-Saturn HD~149026b}

   \author{Federico Biassoni \inst{1,2}
          \and
          Francesco Borsa \inst{2}
          \and
          Francesco Haardt \inst{1,2,3}
          \and
          Monica Rainer \inst{2}
          }

   \institute{DiSAT, Universit\'a degli Studi dell'Insubria, via Valleggio 11,
              I-22100 Como, Italy\\
              \email{fbiassoni@studenti.uninsubria.it}
         \and
             INAF - Osservatorio Astronomico di Brera, via E. Bianchi 46, 
             I-23807 Merate(LC), Italy
            \and
            INFN, Sezione Milano-Bicocca, P.za della Scienza 3, I-20126 Milano, Italy
             }

   \date{submitted Jul 2024}

% \abstract{}{}{}{}{} 
% 5 {} token are mandatory

\abstract{
Advances in modern technologies enable the characterisation of exoplanetary atmospheres, most efficiently exploiting the transmission spectroscopy technique.
We performed visible (VIS) and near infrared (nIR) high-resolution spectroscopic observations of one transit of HD~149026b, a close-in orbit sub Saturn exoplanet.
We first analysed the radial velocity data, refining the value of the projected spin-orbit obliquity. 
Then we performed transmission spectroscopy, looking for absorption signals from the planetary atmosphere. We find no evidence for H$\alpha$, \NaI \, D2$-$D1, \MgI~and \LiI~in the VIS and metastable helium triplet \HeI(2$^3$S) in the nIR using a line-by-line approach. The non-detection of HeI is also supported by theoretical simulations.
With the use of the cross-correlation technique, we do not detect \TiI, \VI, \CrI, \FeI~and VO in the visible, and CH$_4$, CO$_2$, H$_2$O, HCN, NH$_3$, VO in the nIR. Our non-detection of \TiI\ in the planetary atmosphere is in contrast with a previous detection. We performed injection-retrieval tests, finding that our dataset is sensitive to our \TiI \, model. The non-detection supports the \TiI \, cold-trap theory, which is valid for planets with $T_{\rm eq} <$ 2200 K like HD~149026b. 
Even if we do not attribute it directly to the planet, we find a possibly significant \TiI \, signal highly redshifted ($\simeq$+20 \kms) with respect to the planetary restframe. Redshifted signals are also found in the \FeI \, and \CrI \, maps. While we can exclude an eccentric orbit to cause it, we investigated the possibility of material accretion falling onto the star, possibly supported by the presence of strong \LiI \, in the stellar spectrum, without finding conclusive results. The analysis of multiple transits datasets could shed more light on this target.
}


   \maketitle
%
%-------------------------------------------------------------------
\section{Introduction} \label{Introduction}

With advancements in instruments and facilities, the scientific community's focus has shifted from the discovery of exoplanets to their characterization, with one of the main goals the understanding of their atmospheric compositions.
Presently, transmission spectroscopy is the best technique to characterize the properties of the atmospheres of exoplanets \citep[e.g.,][]{Birkby_2018,Madhusudhan_2019}. During transits, exoplanetary atmospheres absorb specific wavelengths of the radiation emitted by their host stars, providing insights into their elemental composition in the stellar light coming to us. High-resolution spectroscopy (resolving power R $\gtrsim$ 20,000) enables us to glean a wealth of information about exoplanet parameters and atmospheric content. This technique allows us to pinpoint atoms and molecules thanks to their unique firm isolating their distinct high cross-section lines. As an example, we are able to identify \NaI \, doublet \citep[e.g.,][]{Wyttenbach_2015}, \CaII \,H\,\&\,K lines \citep[e.g.,][]{Yan_2019}, H$\alpha$ \citep[e.g.,][]{Jensen_2012} and the metastable \HeI\, triplet \citep[e.g.,][]{Allart_2018}. Alternatively, the use of the cross-correlation function (CCF) \citep[e.g.,][]{Snellen_2010,Brogi_2012,Hoeijmakers_2018, Borsa_2022,Prinoth_2023} offers another powerful method to detect species for which strong lines can not be identified.
These types of analyses are mainly employed in the characterization of the atmospheres of the so called hot Jupiters \citep{bell_2018}, i.e., gas giants in close-in orbit. Due to their proximity to the host star, these planets experience intense XUV (10-912 \r{A}) and FUV (912-2,585 \r{A}) irradiation, leading to the heating and possibly inflation of their atmospheres \citep[e.g.,][]{Owen_2019,Biassoni_2023}. 

Among the class of gas giants, a particularly interesting case is that of HD~149026b. The planet was first detected using the radial velocity method by \citet{Sato_2005}, who measured a radius and mass (in Jupiter's units) of $R_p = 0.725 \pm 0.05 \, R_J$, $M_p = 0.36 \pm 0.02 \, M_J$, respectively. HD~149026b is then more massive but smaller in size than Saturn, despite receiving intense irradiation from its parent star, a fact that should lead to an inflated atmosphere \citep{Sato_2005}. The planet's high density, such as $\rho_p=1.18_{-0.30}^{+0.38}$\,\,g cm$^{-3}$ \citep{Wolf_2007}, $\rho_p=1.59_{-0.36}^{+0.38}$\,\,g cm$^{-3}$ \citep{Torres_2008}, $\rho_p=2.1 \pm 0.8$\,\,g cm$^{-3}$ \citep{Southworth_2010}, suggests a core rich in metals, as proposed by \citet{Sato_2005}, who estimated a core mass $\approx 67 \, M_{\oplus}$, with $M_{\oplus}$ is the Earth's mass. Subsequent studies echoed these high core estimations.  \citet{Fortney_2006} claimed a significant presence of heavy elements in both the core and envelope, estimating values $\approx 60 - 93 \, M_{\oplus}$; \citet{Burrows_2007} inferred a core mass ranging between $80 - 110 \, M_{\oplus}$; \citet{Baraffe_2008} and \citet{Carter_2009} estimated a heavy element core mass of $\approx 60 - 80 \, M_{\oplus}$ and $\approx 45 - 70 \, M_{\oplus}$, respectively.

A combination of both a high core mass and high stellar metallicity \citep[{{${[Fe/H]}= 0.36 \pm 0.05$,}}][]{Sato_2005} appears to be crucial prerequisites for a large atmospheric metallicity of the planet \citep{Zhang_2018}.
Several atmospheric models have been proposed to unravel the composition of HD~149026b. \citet{Fortney_2006} observed a hot stratosphere attributed to the absorption of stellar flux by TiO and VO, while \citet{Stevenson_2012}, analysing Spitzer secondary eclipse observations, adopted a chemical equilibrium model and suggested the presence of large amounts of CO and CO$_2$, with moderate heat redistribution, high metallicity, and absence of thermal inversion on the dayside. \citet{Zhang_2018} analysed Spitzer phase curves observations at 3.6 and 4.5 $\mu$m and inferred a high albedo, possibly explained by the presence of reflective cloud layers in the planet's upper atmosphere. Recently analyses based on transmission spectroscopy conducted with the High Dispersion Spectrograph \citep{Noguchi_2002} on the Subaru telescope were performed by \citet{Ishizuka_2021}. They reported a tentative $\simeq 4.4 \sigma$ detection of \TiI, and a marginal $\simeq 2.8 \sigma$ detection of \FeI, alongside non detections of \ScI, \VI, \CrI, \MnI, \CoI\, and TiO. The presence of \TiI\, without TiO suggests a supersolar C/O ratio \citep{Ishizuka_2021}.
Recently, \citet{Bean_2023} analysed the HD~149026b dayside emission obtained with the James Webb Space Telescope (JWST). Similarly to prior studies, they concluded that the planet exhibits a highly super-solar metallicity ($[M/H] = 2.09^{+0.35}_{-0.32}$) and a high carbon-to-oxygen ratio ($[C/O] = 0.84 \pm 0.03$). These results further support the detection of CO$_2$ and H$_2$O in HD~149026b. 
The same dataset was analysed by \citet{Gagnebin_2024} using a 1D radiative-convective-thermochemical equilibrium models from which new values of planet atmospheric metallicity are estimated to be $\simeq 1.15$ if VO isn't included in the models, and $\simeq 1.30$ if it is. These values are $\simeq$ 10 times smaller than the results of \citet{Bean_2023}. 

\citet{Spinelli_2023} updated the HD~149026b parameters by exploiting more precise GAIA DR2 \citep{Gaia_2018} distances, inferring a mass and radius of $M_p = 0.28 \pm 0.03 \, M_J$ and $R_p = 0.74 \pm 0.02 \, R_J$, respectively. These values yield a planetary density of $\rho_p = 0.86 \pm 0.09$\,\, g cm$^{-3}$, consistent with the lowest estimates of previous works, e.g., $\rho_p=0.750^{+0.089}_{-0.077}$\,g cm$^{-3}$ as in \citet{Bonomo_2017}, and $\rho_p = 0.85^{+0.10}_{-0.09}$\,g cm$^{-3}$ as in \citet{Carter_2009}.
Given the results by \citet{Spinelli_2023}, we decided to shed more light on the atmospheric content of HD~149026b with high-resolution transmission spectroscopy. To this aim, we collected data from two transits at the Telescopio Nazionale Galileo (TNG) equipped with high-resolution visible and near-infrared spectrographs, i.e., the High Accuracy Radial velocity Planet Searcher North (HARPS-N) (R $\simeq$ 115,000) and GIANO-B (R $\simeq$ 50,000), respectively. 

In this work, we describe our observations, data analysis and scientific results. The paper is organized as follows: details regarding the observations can be found in \S~\ref{Observations}, while \S~\ref{RM_model} and \S~\ref{Transmission_spectroscopy} cover the modelling of Rossiter-McLaughlin effect, center-to-limb variations and data analyses. Discussion of the results of the data analysis and concluding remarks are detailed in the final \S~\ref{Discussion}.

%--------------------------------------------------------------------
\begin{figure*}
\center
\includegraphics[width=0.48\textwidth]{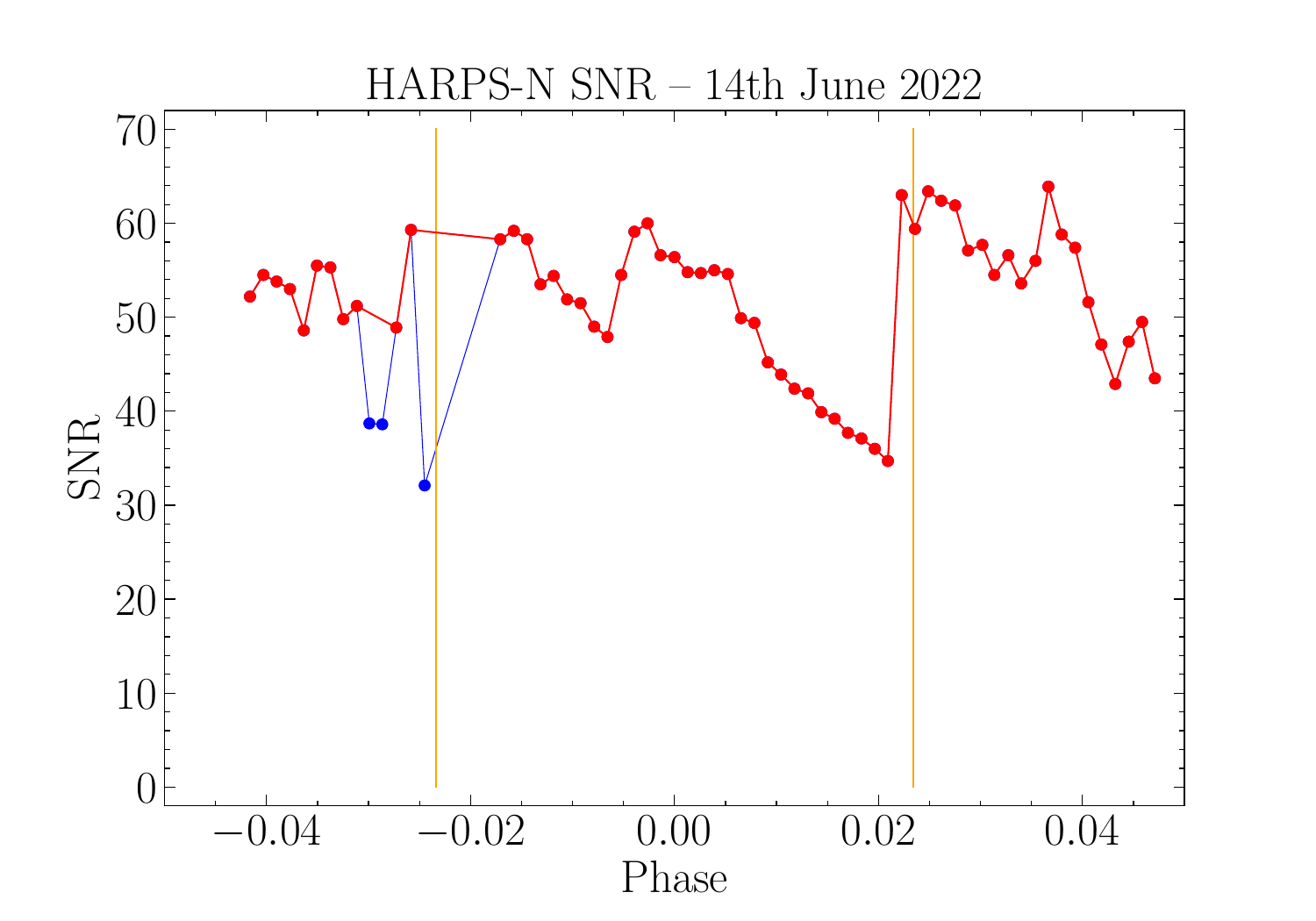} 
\includegraphics[width=0.48\textwidth]{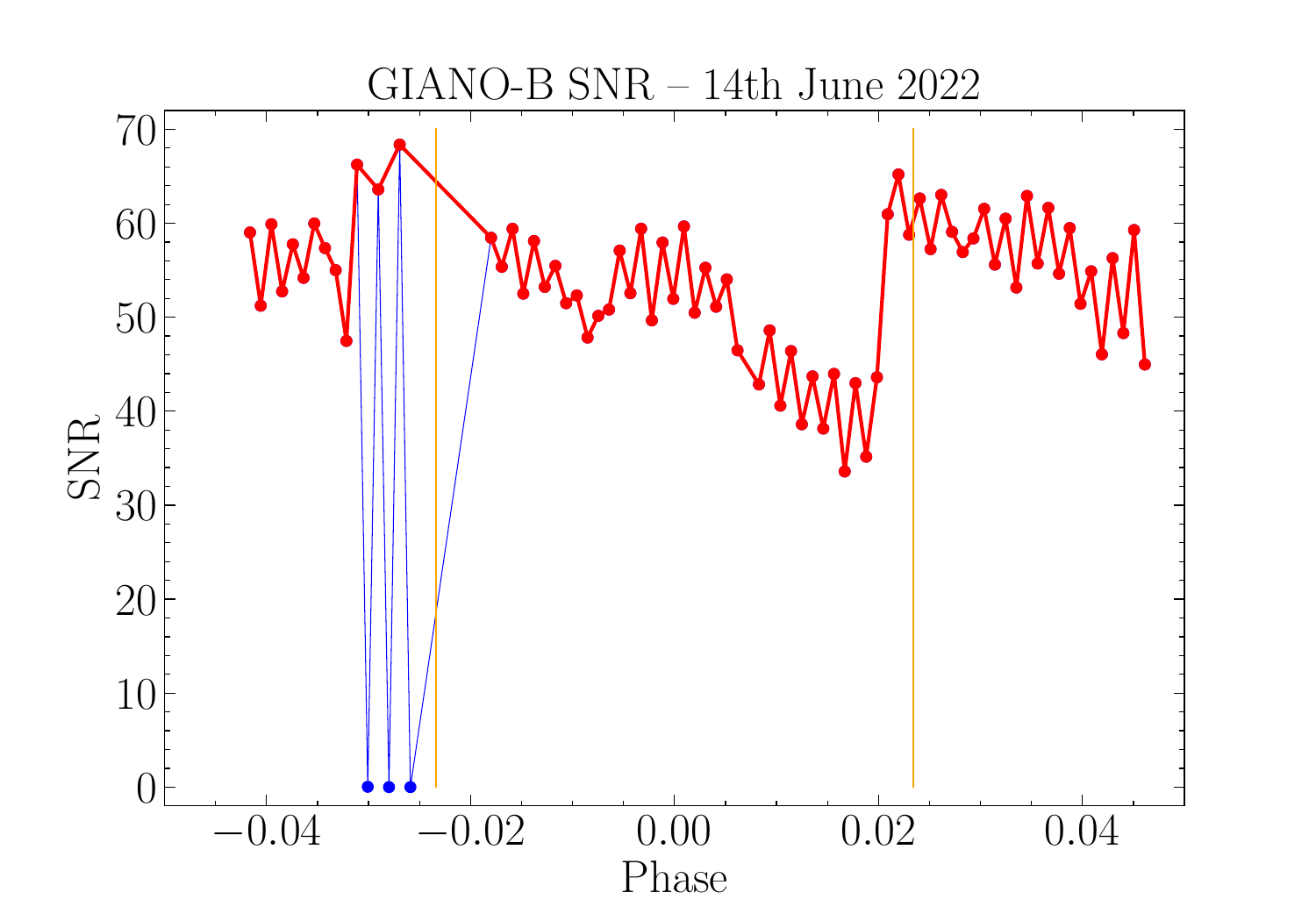}
\caption{HARPS-N and GIANO-B SNR for the second night evaluated at 5,500 and 16,300~\AA{}, respectively. The red dots represent the SNR of the spectra used for transmission spectroscopy analysis, while the blue ones are discarded. The two vertical orange lines denote the T$_1$ and T$_4$ transit contact points.}
\label{fig:harpsn_gianob_snr}
\end{figure*}

\section{Observations} \label{Observations}

We observed one transit of HD~149026b on June 14th, 2022 with the HARPS-N and GIANO-B high-resolution spectrographs, mounted at the Telescopio Nazionale Galileo (Program A45TAC\_30, PI Borsa). We exploited the GIARPS configuration \citep{Claudi_2016}, which allows us to observe simultaneously with the two spectrographs. A second scheduled transit on 3rd April 2022 was lost due to bad weather.
The wavelength range of the acquired spectra span from visible (VIS, 3,900 - 6,900~\AA{}, HARPS-N) to near-infrared (nIR, 9,400 - 24,200~\AA{}, GIANO-B).
With GIANO-B, we collected 78 spectra with exposures of 200 sec per nodding position (ABAB pattern). Meanwhile with HARPS-N we observed a total of 64 spectra with an exposure time of 300 sec. All the spectra were reduced with the standard DRS pipelines v3.7 and v1.6.1 for HARPS-N \citep{Cosentino_2012} and GIANO-B (\citet{Rainer_2018,Harutyunyan_2018}), respectively.
A summary of the observations is shown in Table \ref{table:1}.

During the observations, we lost about half an hour at the transit ingress because of a problem with the instrument guiding system. A strong loss of flux was experienced also during the transit, which was solved by performing a repointing procedure. 
Figure \ref{fig:harpsn_gianob_snr} shows the signal-to-noise ratio (SNR) obtained with HARPS-N and GIANO-B at 5,500 and 16,300~\AA{}, respectively.
For our analysis, we discarded all the spectra with low SNR with respect to the adjacent ones (see Fig.~\ref{fig:harpsn_gianob_snr}), ultimately performing our analysis on a total of 61 (75) HARPS-N (GIANO-B) spectra.

\begin{table}
\centering
\renewcommand{\arraystretch}{1.3}
% \fontsize{11}{12}\selectfont
\caption{log of the observations performed with the two instruments$^{(a)}$.}
\label{table:1} 
\begin{tabular}{ccc}
    	% P{2.3cm}%
    	% P{2.3cm}% 
    	% P{2.3cm}% 
    	% }
	    \hline 
    	\hline

    	Instrument                  &% 
    	\# spectra  (Out/In)   &% 
            <SNR>      \\ 
    	%---------------
    	\cline{1-3}
    	HARPS-N     &
    	61 (30/31)  &
    	52.0        \\
    	%--------------
    	GIANO-B     &
    	75 (37/38)  &
    	53.6        \\
    	%--------------
    	\hline
    	\hline
	\end{tabular}
\footnotesize
\begin{flushleft}
(a) Parenthesis specify the spectra collected in and out of transit. The average SNR is taken at 5,500 \AA{} and 16,300 \AA{} for HARPS-N and GIANO-B respectively.
\end{flushleft}
\end{table}

%--------------------------------------------------------------------
%\section{Data Analysis} \label{Data analysis}

\begin{table*}
\centering
\renewcommand{\arraystretch}{1.0}
\fontsize{10}{13.5}\selectfont
\caption{physical parameters for the HD 149026 system.}
\label{table:2}
\begin{tabular}{
m{5.0cm}
P{5.0cm}
P{5.0cm}
}
\hline
\hline
% -------------------------
Parameter       &
Symbol   &
Value [Unit]    \\
% -------------------------
\hline
Stellar Parameters  &
                    &
                    \\
    % -------------------------
    \hline
            
    Age $^{(a)}$    &
             &
    $2.60 \pm 0.20 \,\, $[Gyr] \\
    
    Effective Temperature $^{(g)}$  &
    $T_{eff}$                       &
    $6075  \,\, $[K]         \\ %.75_{-74.73}^{+107.25} \,\, $[K]         \\
    
    Spectral Class $^{(c)}$     &
    &
    G0 \\
    
    Stellar Mass $^{(h)}$           &
    $M_{\star}$                         &
    $1.09 \pm 0.13 \,\, $[M$_{\odot}$] \\
    
    Stellar Radius $^{(h)}$          &
    $R_{\star}$                          &
    $1.47 \,\, $[R$_{\odot}$] \\
    
    Projected Rotation Speed $^{(f)}$  &
    $v \sin{i}$                         &
    $6.10_{-0.48}^{+0.47} \,\, $[km s$^{-1}$] \\
    
    Surface Gravity  $^{(b)}$   &
    $\log{g}$                   &
    $4.37 \pm 0.04 \,\, $[cm s$^{-2}$]  \\

    Metallicity $^{(a)}$        &
    $[Fe/H]$                    &
    $0.36 \pm 0.08 \,\,$ [dex]\\

% -------------------------
\hline
Planetary Parameters    &
                        &
                        \\
% -------------------------
    \hline
    Planet Mass $^{(h)}$        &
    $M_{p}$                 &
    $0.28 \pm 0.03 \,\, $[M$_{\rm J}$] \\
    
    Planet Radius $^{(h)}$      &
    $R_{p}$ &
    $0.74 \pm 0.02 \,\, $[R$_{\rm J}$] \\
    
    Equilibrium Temperature $^{(d)}$    &
    $T_{eq}$    &
    $1634^{+90}_{-23} \,\, $[K] \\
    
    Planet Density $^{(h)}$ &
    $\rho_{p}$              &
    $0.86 \pm 0.09 \,\, $[g cm$^{-3}$] \\
    
    % Surface Gravity $^{(a)}$    &
    % $log{g_p}$                  &
    % $3.085^{+0.035}_{-0.034} \,\, [cm/s^2]$ \\
        
% -------------------------
\hline
Orbital Parameters      &
                        &
                        \\
    % -------------------------
    \hline
    Epoch $^{(f)}$  &
    $T_0$   &
    $2454597.702816162_{-0.004317}^{+0.003742} \,\, $[BJD]\\
    
    Period $^{(a)}$  &
    $P$              &
    $2.8758916 \pm 0.0000014 \,\, $[days]\\
    
    Transit Duration $^{(e)}$   &
    $T_{14}$                    &
    $3.23 \pm 0.15 \,\, $[h]  \\
    
    Systemic Velocity $^{(f)}$  &
    $\gamma$                    &
    $-18.0319_{-0.0008}^{+0.0008} \,\, $[km s$^{-1}$]\\
    
    Radial Velocity Amplitude $^{(f)}$  &
    $K_{\star}$                             &
    $38.3 \pm 3.1 \,\, $[m s$^{-1}$]     \\
    
    Semi-major Axis $^{(h)}$    &
    $a$                         &
    $0.041 \pm 0.002 \,\,$ [AU]\\
    
    Inclination $^{(f)}$                &
    $i$                                 &
    $84.1130^{+0.5259}_{-0.4986} \,\, $[deg]   \\
    
    Eccentricy $^{(a)}$ &
    $e$                 &
    $< 0.013 \,\, $     \\
    
    Projected Obliquity $^{(f)}$ &
    $\lambda$                    &
    $2.5040_{-5.3317}^{+5.3640} \,\, $[deg]\\

\hline
\hline

\multicolumn{3}{l}{
(a) \citet{Bonomo_2017};
(b) \citet{Stassun_2017};
(c) \citet{Ment_2019};
(d) \citet{Torres_2008}}\\
\multicolumn{3}{l}{
(e) \citet{Albrecht_2012};
(f) This work;
(g) \citet{Gaia_2018};
(h) \citet{Spinelli_2023}}\\
    \end{tabular}
\end{table*}

\section{Rossiter-McLaughlin effect} \label{RM_model}

During a transit, the partial occultation of the stellar disk by the planet induces a distortion in the observed stellar spectrum, leading to apparent variations of the derived stellar radial velocities (RVs). This specific phenomenon is commonly referred to as the Rossiter-McLaughlin (RML) effect \citep{Rossiter_1924,McLaughling_1924}. The amplitude and shape of the RML effect are mainly dependent on the projected stellar rotation velocity ($v \sin{i}$), projected spin-orbit inclination, impact parameter and $R_p/R_\star$ ratio, where $R_\star$ is the stellar radius. An analysis of the RML effect enables the extraction of these planetary and stellar parameters which are later used in the transmission spectroscopy (Sect. \ref{single_line_analyses}). In order to model the RML effect, we employed the {\fontfamily{pcr}\selectfont CaRM} code, a semi-automated tool \citep{Cristo_2022} that employs an MCMC algorithm. Details of our analysis of the RML effect are presented in Appendix~\ref{appendice}.
Globally, the values of projected rotational speed, systemic velocity, radial velocity amplitude and inclination we obtain are in agreement with literature results.
In particular, we refine the projected spin-orbit angle $\lambda = 2.5 \pm 5.3$ deg, in agreement with the literature value \citep[$\lambda=12 \pm 7$ deg,][]{Albrecht_2012} but with smaller error-bars.

\begin{figure}
% \center
\includegraphics[width=0.5\textwidth]{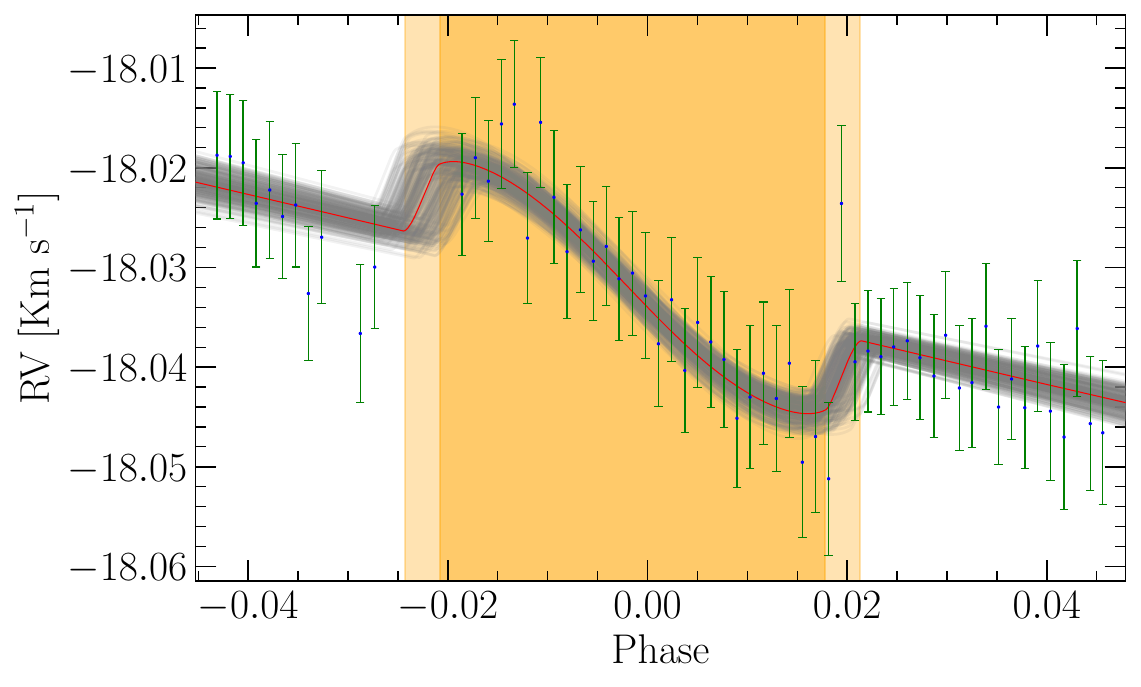} 
\caption{Result from the RML fitting procedure performed with {\fontfamily{pcr}\selectfont CaRM}. The red line is the best fit model, the grey shadows are a random sample of the posterior distribution. The yellow regions show the ingress and egress of the transit, while the orange region shows where the planet is fully transiting the stellar disk.}
\label{fig:RM_fit}
\end{figure}

%%%%%%%%%%%%%%%%%%%%%%%%%%%%%%%%%%%%%%%%%%%%
\section{Transmission spectroscopy}\label{Transmission_spectroscopy}

The main goal of our observations is to look for the atmosphere of HD~149026b using high-resolution transmission spectroscopy. We started our analysis from the merged 1D spectra (s1d) provided by the HARPS-N DRS. Similarly, we analysed the 71$^{th}$ GIANO-B spectral order searching for the presence of the metastable helium triplet \HeI(2$^3$S) at $\simeq 10,830$~\AA{}. We chose to work with the echelle spectral orders (ms1d) in the nIR band instead of the merged s1d spectra to improve the normalisation of the GIANO-B spectra.

\subsection{Telluric correction and wavelength calibration}
\begin{figure*}
\centering
\includegraphics[width=1\textwidth]{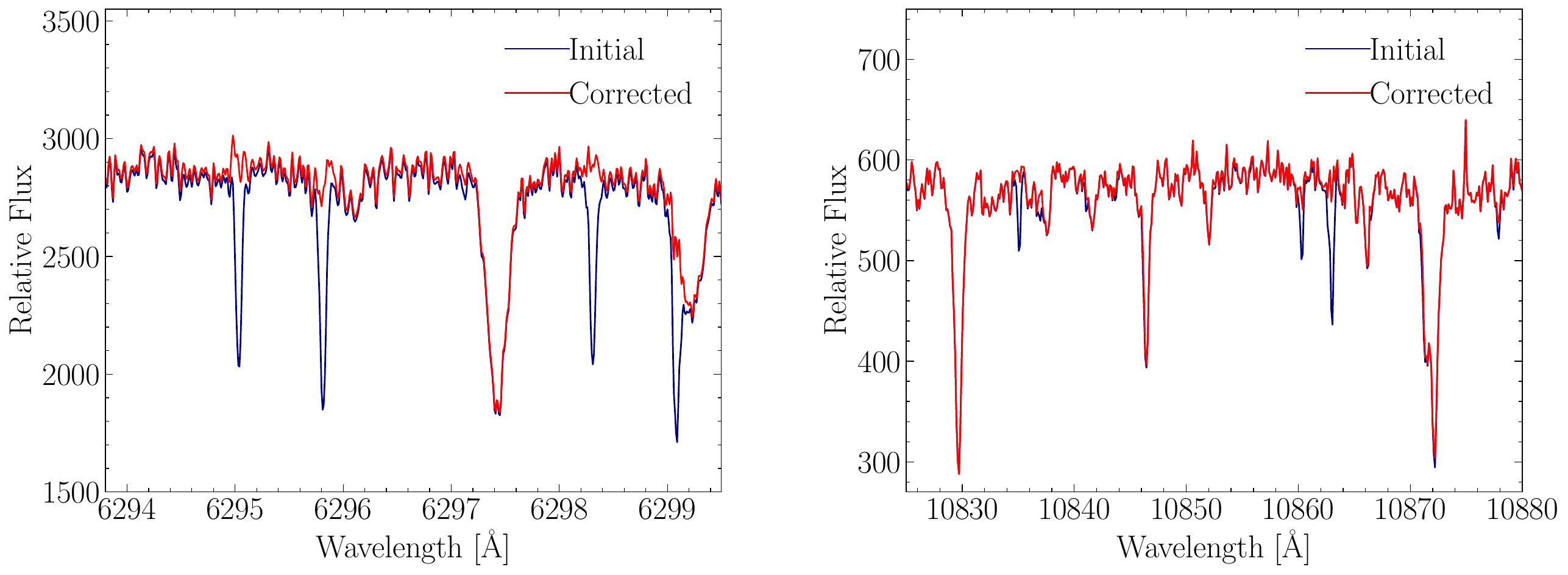}
\caption{Example of {\fontfamily{pcr}\selectfont Molecfit} telluric correction. The left figure refers to the HARPS-N telluric correction in proximity of the H$\alpha$ line, while right figure to the GIANO-B telluric correction in proximity of the \HeI(2$^3$S). Both spectra are taken from the first exposure.}
\label{fig:telluric}
\end{figure*}

We performed correction for telluric H$_2$O and O$_2$ lines on all HARPS-N spectra using {\fontfamily{pcr}\selectfont Molecfit} v. 4.3.1 \citep{Smette_2015,Kausch_2015}, following the guidelines specified in \citet{Allart_2017} but slightly modifying the wavelength ranges of the correction. The left panel of Fig. \ref{fig:telluric} provides an example of telluric correction obtained using {\fontfamily{pcr}\selectfont Molecfit}, near the H$\alpha$ line.

We used {\fontfamily{pcr}\selectfont Molecfit} v. 4.3.1 also to correct the GIANO-B spectra, but following a different recipe. The GIANO-B wavelength calibration may be slightly inconsistent between the different echelle orders due to the characteristics of the U-Ne calibration lamp (the number of useful emission lines varies greatly between the orders). While {\fontfamily{pcr}\selectfont Molecfit} accounts for some wavelength shift in the observed spectrum, trying to correct the whole merged GIANO-B spectra resulted in p-Cygni-like residuals due to the slight misalignment of the echelle orders. For this reason we worked with the ms1d GIANO-B spectra, where the echelle orders are still separated, and we ran {\fontfamily{pcr}\selectfont Molecfit} on each order independently. Another complication arose from the fact that the nIR wavelength range is heavily affected by the telluric absorption: due to the combination of the wide wavelength range of GIANO-B and the large telluric contribution, the computational time needed to model and remove the telluric lines from all the 50 orders of a single GIANO-B spectrum is quite high. In order to speed up the process, we computed the model only once every ten spectra, assuming that the atmospheric conditions did not vary too much in this time. To account for any possible variation due to the position of the star on the slit (A and B position of the nodding observational strategy), we worked separately on the A and B spectra, computing one model for every ten A spectra and one for every ten B spectra. While we globally removed H$_2$O, O$_2$, CO$_2$, N$_2$O, CH$_4$ and CO from the whole spectrum, working on the separate orders allowed us to model only the atmospheric molecules present in each echelle order. Right panel of Fig.~\ref{fig:telluric} provides an example of telluric correction near the \HeI(2$^3$S) lines.

\subsection{Transmission spectrum extraction}

We extracted the transmission spectrum following the guidelines outlined in \citet{Wyttenbach_2015}, performing the procedure in a short wavelength interval bracketing each of the atmospheric lines investigated (see Table \ref{table:range}).
As a first step, all the spectra were normalized and shifted to the stellar reference frame. Then, we generated an average reference stellar spectrum (Master-Out, $M_{\rm out}$), by performing a weighted mean of all the out-of-transit spectra. 
Each spectrum was then divided by $M_{\rm out}$ and the residual spectra $R_i$ derived accordingly. These $R_i$ were further normalized to eliminate any trend possibly given by imperfect continuum atmospheric dispersion corrections.
In order to maximize any potential planetary signal (if present), all $R_i$ were shifted into the planet reference frame, considering a circular orbit with parameters form Table \ref{table:2}. Then we computed a weighted mean of all the in-transit residual spectra to obtain the final average transmission spectrum. These transmitted spectra have shorten wavelength range with respect to the initial intervals chosen for each element analysed; this is due to the Doppler shifts and subsequent interpolation procedures adopted during the transmission spectroscopy analysis that bring divergence at the edges of the wavelength ranges (Table \ref{table:range}). After each interpolation the wavelength bin width was maintained constant at the value of 0.01~\AA{}.

\begin{table}
\centering
\renewcommand{\arraystretch}{1.0}
\fontsize{10}{13}\selectfont
\caption{interval used for the transmission spectroscopy analysis.}
\label{table:range} 
\begin{tabular}{ccc}
    	% P{2.3cm}%
    	% P{2.3cm}% 
    	% P{2.3cm}% 
    	% }
	    \hline 
    	\hline

    	Element         & 
    	Initial $\lambda$ range [\AA{}]& 
    	Final $\lambda$ range [\AA{}] \\ 
    	%---------------
    	\cline{1-3}
    	MgI     &
    	5,125-5,195  &
    	5,140-5,190  \\
    	%--------------
    	NaI D2-D1     &
    	5,850-5,930  &
    	5,885-5,905 \\
            %--------------
    	H$\alpha$     &
    	6,525-6,595  &
    	6,550-6,580  \\
            %--------------
    	LiI     &
    	6,670-6,720  &
    	6,700-6,715  \\
            %--------------
    	HeI(2$^3$S)     &
    	10,800-10,880  &
    	10,825-10,840  \\
    	%--------------
    	\hline
    	\hline
	\end{tabular}
\end{table} 

\begin{figure*}
\centering 
\includegraphics[width=1.0\textwidth]{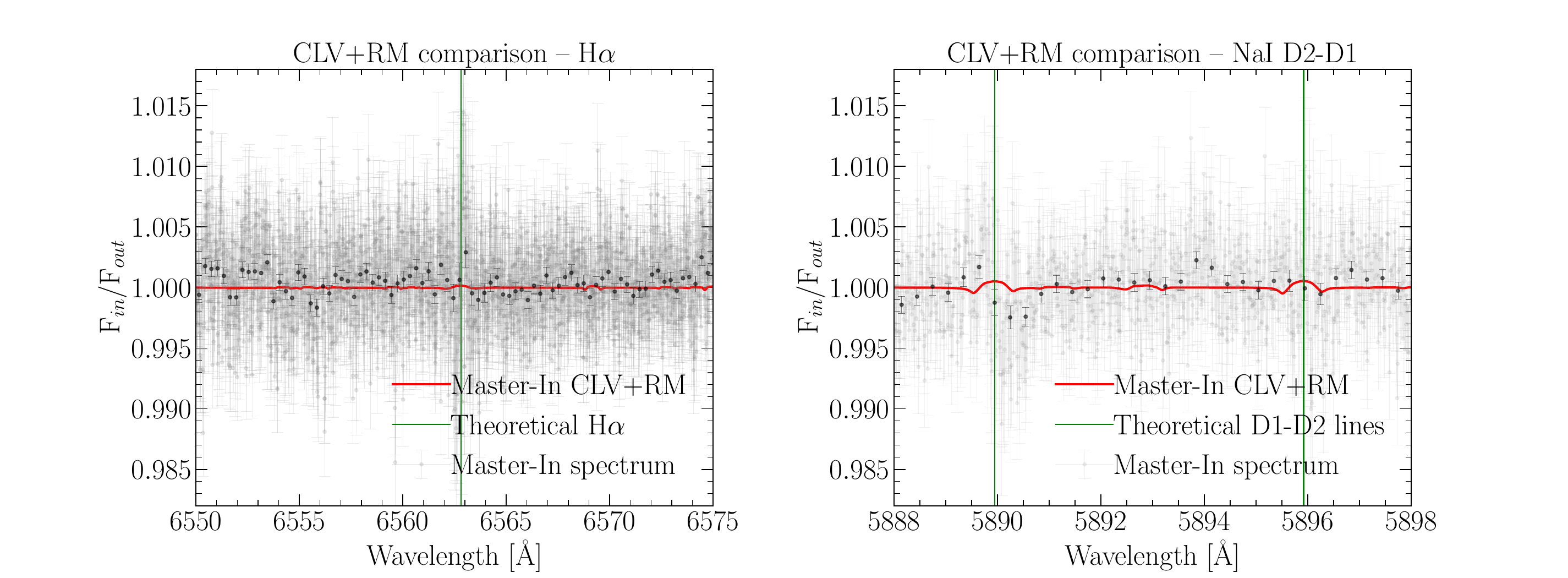} 
\caption{CLV+RM models (red) compared with transmission spectra (light grey) around the H$\alpha$ and \NaI \, D doublet wavelength regions. Green lines show the position of the theoretical values of H$\alpha$, \NaI \, D2 and D1 lines. The CLV+RML model contamination is well within the data noise.}
\label{fig:RM_Data_comparison}
\end{figure*}

\subsection{Single lines analysis}\label{single_line_analyses}

We extracted the transmission spectrum in the wavelength ranges of \MgI\, triplet (5,167.32 - 5,172.68 - 5,183.60~\AA{}), \NaI\, D2-D1 (5,889.95 - 5,895.92~\AA{}), H$\alpha$ (6,562.81~\AA{}), \LiI\, (6,707.76~\AA{}) in the optical band, and of the \HeI(2$^3$S) triplet (10,829.09 - 10,830.25 - 10,830.34~\AA{}) in the infrared band\footnote{Wavelengths are given in air.}.

During a transit, one must consider that the host star is not an homogeneously bright disc, rather its surface brightness changes with the the distance from disc centre (centre-to-limb variations, CLV). Moreover, the star rotates. 
These stellar properties affect the emitted spectrum occulted during transit through CLV and RML effect, respectively, that, in turn can affect the transmission spectrum in various ways, e.g., by modifying the shape of line profiles and/or by causing false atmospheric detections \citep[e.g.,][]{Yan_2017,Borsa_2018,Casasayas_2020}.  
Given our target's characteristics of low projected rotational velocity and small $R_p/R_{\star}$ ratio, we expect minimal-to-low contamination from RML and CLV. In order to verify if this is indeed the case, we modelled CLV and RML as outlined in \citet{Yan_2017} on few lines (H$\alpha$, \NaI\, doublet and \MgI\, triplet), using the same methodology employed in \citet{Borsa_2021}. 
The star was modelled as a disc, and mapped on fine grid with 0.01~$R_\star$ resolution. At each point of the grid we calculated the angle between the normal to the stellar surface and the line of sight, $\mu=\cos{\theta}$, and the projected rotational velocity (rescaling the $v \sin{i}$ and using rigid body rotation).
A spectrum was then assigned to each point of the grid by a quadratic interpolation on $\mu$, and Doppler-shifting according to the stellar rotation the model spectra created using the tool Spectroscopy Made Easy \citep[SME,][]{Piskunov_2017}, with the line list from the VALD database \citep{Ryabchikova_2015} and ATLAS9 \citep{Kurucz_1993, Castelli_2003} stellar atmospheric models. The model spectra were created with null rotational velocity for 21 different $\mu$ values at the resolving power of HARPS-N.
Using the orbital information from Table~\ref{table:2}, we then simulated the transit of the planet, calculating the stellar spectrum for different orbital phases as the average spectrum of the non-occulted modelled sections. In the last step, we divided each spectrum for a master stellar spectrum calculated out of transit, obtaining the information of the relevance of CLV+RML effects at each in-transit orbital phase.
We then moved everything in the planetary rest frame, calculating the simulated CLV+RML effects on the transmission spectrum. 

\begin{figure*}[ht!]
\centering 
\includegraphics[width=1\textwidth]{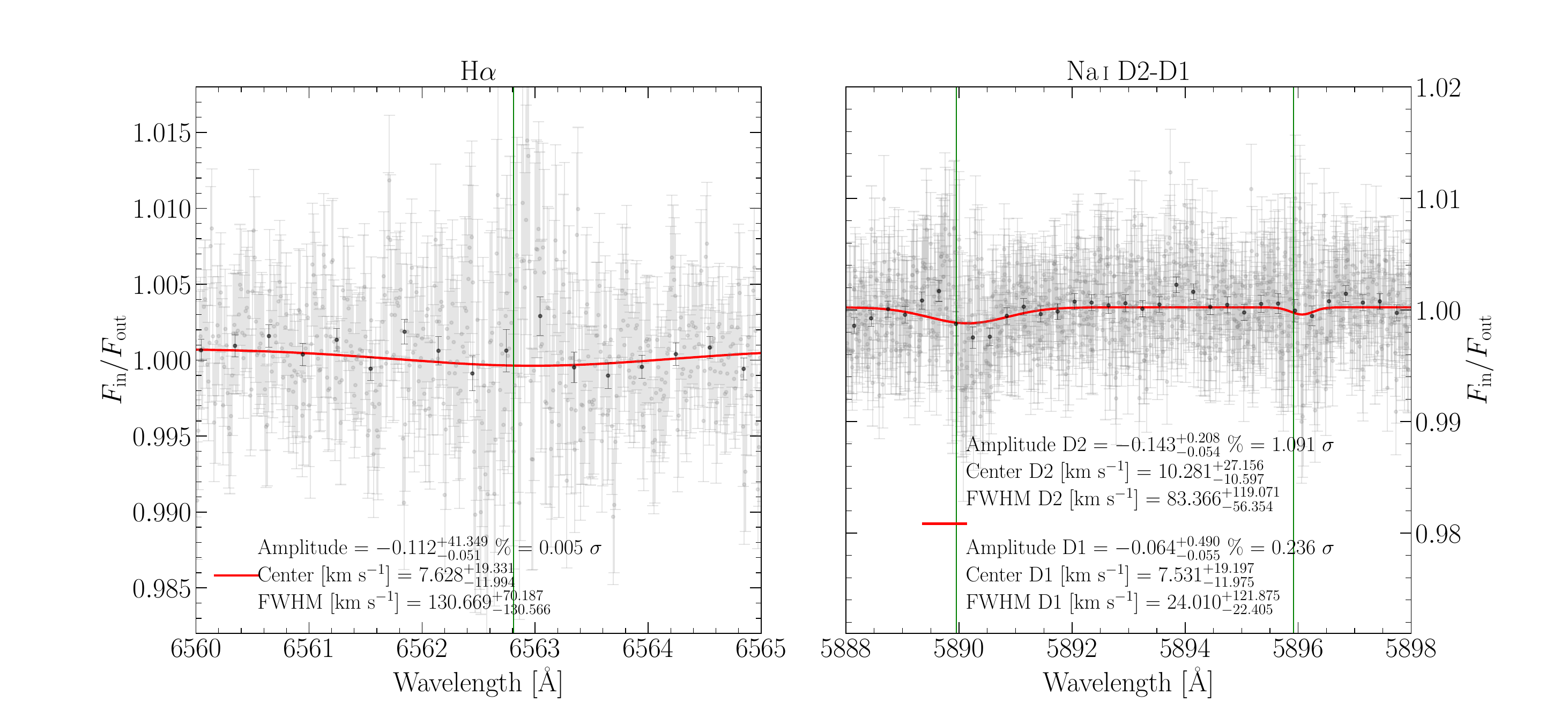}
\caption{H$\alpha$ and \NaI \,D2-D1 transmission spectra. The grey points are our transmission spectra, the red line the best Gaussian fit while the black dots are a resample of the transmission spectra with bin size of 0.3\,\AA{}.
}
\label{fig:Fit}
\end{figure*}

In Fig.~\ref{fig:RM_Data_comparison} we show the comparison between the transmission spectra of H$\alpha$ and \NaI \, D doublet and the simulated CLV+RML contamination. The latter is well within the noise of the data, confirming expectations. Since CLV+RML are negligible in the transmission spectrum, we ignore them in our subsequent analyses.

We performed a Gaussian fit around all the lines of interest, using a linear regression method (see footnote in Appendix~\ref{appendice}). 
In the visible band, our analysis led to non-detection for all the considered lines, i.e., H$\alpha$, \NaI\, D2-D1, \MgI\, triplet and \LiI\,.
Examples of the results are shown in Fig.~\ref{fig:Fit}. 
We estimated upper limits of such lines by calculating the standard deviation within $\pm 0.5$~\AA{} from their reference wavelengths (see Table~\ref{table:4}). 

Indeed, the depth of stellar lines strongly affects our ability to put strong limits on the atmospheric heights. We used $M_{\rm out}$ to measure the stellar flux intensity within the core of all these lines relative to the continuum. 
As shown in Table~\ref{table:4}, most of the transmission signals we were looking for are affected by a stellar flux which is quite low.
As a result, the SNR of the transmission spectrum of the planet is constrained by such low values of the flux, potentially making any possible planetary signal lost in the noise. 

We also surveyed the nIR band in search of the metastable helium triplet \HeI(2$^3$S) at $\simeq 10,830$~\AA{}. This state of Helium predominantly appears in planets orbiting K-type stars \citep{Oklopčić_2019}, and its abundance, and relative absorption signal, depends upon various factor, e.g., the hardness ratio ($L_{\rm XUV}/L_{\rm FUV}$) of the stellar spectrum, the planetary orbital distance, the planetary Hill's radius, and the He-to-H abundance ratio in the planet's outer atmosphere \citep[e.g.,][]{Biassoni_2023}. 

\begin{figure*}
\centering
\includegraphics[width=1\textwidth]{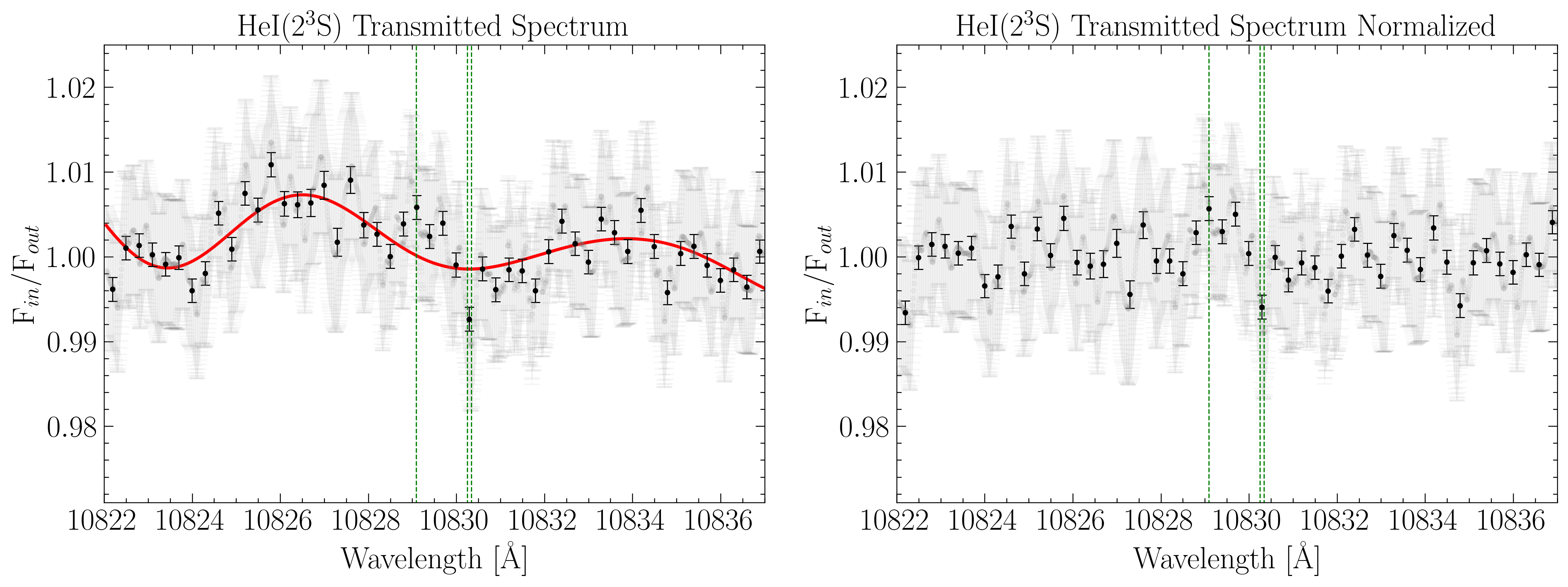}
\caption{\HeI(2$^3$S) transmission spectrum. Left panel: in grey $F_{\rm in}/F_{\rm out}$ obtained with the transmission spectroscopy analysis. A sinusoidal shape is evident. Red line is the sinusoidal fit obtained using the relation described in the text. Right panel: the grey transmission spectrum divided by the sinusoidal red fit. Black dots represent the binned spectra with equally spaced bins of 0.3~\AA{}, while vertical dashed green lines show the position of the theoretical \HeI(2$^3$S) lines in vacuum.}
\label{fig:HeI3_normalization}
\end{figure*}

We applied the same technique used for the visible band to extract the transmission spectrum close to the \HeI(2$^3$S) lines. 
The result shows substantial noise and a sinusoidal pattern (Fig. \ref{fig:HeI3_normalization}), which is a known artefact in GIANO-B transmission spectra \citep[e.g.][]{Guilluy_2023}. In order to improve the normalization, we fitted a sinusoidal function of the form $A\,\cdot\, \sin(x) \,+\, B\,\cdot\, \cos(b x)\, +\, mx \,+\, C$, for which we divided the transmission spectrum (grey curve, Fig. \ref{fig:HeI3_normalization}). 
During the analysis, we took into account the two telluric OH doublets (10,829.46-10,829.15~\AA{} and 10,831.38-10,831.29~\AA{}, wavelengths in air, \citet{Oliva_2013}), whose emission can notably affect this wavelength region \citep[e.g.,][]{Guilluy_2023}. We shifted them in the stellar reference system after correcting for the systemic velocity and the barycentric velocity of the Earth. In this reference system these lines fall at 10,829.84-10829.54~\AA{} and 10,831.77-10,831.67~\AA{} respectively. To remove them, we masked completely the regions between [10,832.38 - 10,832.89]~\AA{} and [10,834.53 - 10,834.81]~\AA{}.

\begin{figure*}
\sidecaption
\begin{minipage}{12cm}
  \includegraphics[width=\textwidth]{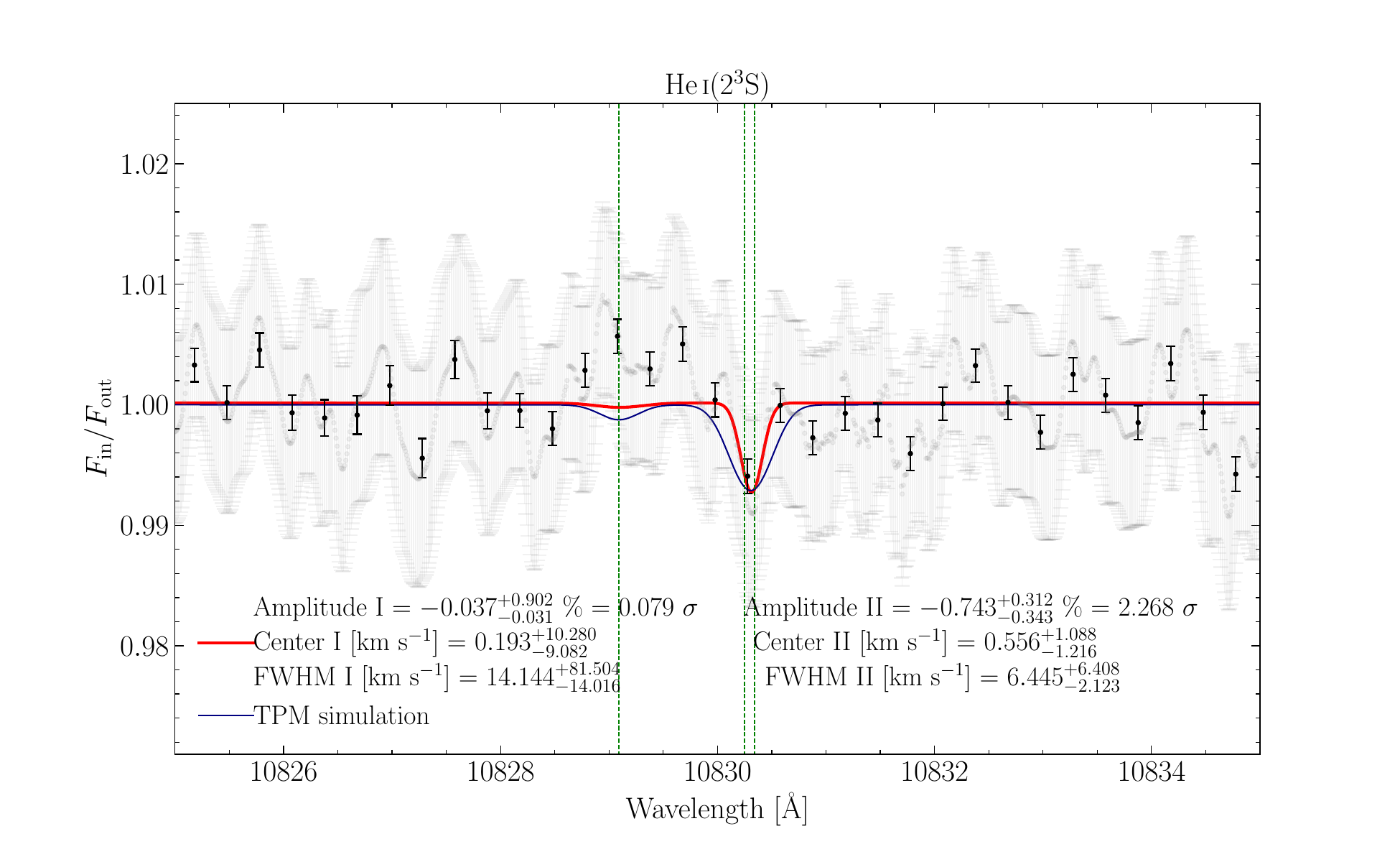}
\end{minipage}
     \caption{\HeI(2$^3$S) normalized transmission spectrum in comparison with the double Gaussian fit and TPM simulation.} \label{fig:HeI3_fit}
\end{figure*}

Similarly to the analysis in the visible band, the helium profile was fitted with two Gaussian profiles like the \NaI\, doublet. The doublet lines at $10,830.25 - 10,830.34$~\AA{} were considered blended \citep{Kirk_2022}. The prior on the center of the doublet, $\mu = 10,830.29$~\AA{} was set between $\pm 0.2$~\AA{} to prevent significant deviation from the fit due to the high noise level in the spectrum. The fit hinted at an absorption pattern near the \HeI(2$^3$S) doublet (figure \ref{fig:HeI3_fit}), but the profile lacked a clear definition, and its deviation from the continuum was only at $\simeq 2.3 \, \sigma$. 

The absence of a significant \HeI(2$^3$S) signal in the planetary atmosphere was further investigated theoretically.
Using the 1D photo-ionization hydrodynamic code ATES \citep{Caldiroli_2021} and the Transmission Probability Module (TPM) \citep{Biassoni_2023}, we modelled the outflow of HD~149026b, estimating its mass loss rate alongside the number density of \HeI(2$^3$S) and the transmission spectrum. To achieve this simulation, we employed the HD~149026 spectrum provided by \citet{Behr_2023} rescaled at the planetary orbital distance assuming a circular orbit.
In the simulation we used planetary and stellar parameters from Table \ref{table:2}. To match our observations, we convolved the simulated absorption profile with the GIANO-B instrumental resolution (R = 50,000, assumed as Gaussian), considering also the planet's rotation under the assumptions of tidal locking and a radius equal to $R_{\rm eff} = R_p \sqrt{1 + h/\delta}$, where $h$ represents the theoretical absorption depth at the \HeI(2$^3$S) doublet core derived from TPM, and $\delta = (R_p/R_{\star})^2$ denotes the transit depth.
The simulation produces a mass-loss rate of $\approx 5\times 10^{10}$\,g\,s$^{-1}$, resulting in an absorption depth of \HeI(2$^3$S) of $\simeq 0.7 \%$ (blue curve in Figure \ref{fig:HeI3_fit}), of the same order of the error-bars in the data.

% \begin{figure*}
% \centering
% \includegraphics[width=1\textwidth]{Figure/HeI3_result_paper_HD149_AIR_v2.pdf}
% \caption{\HeI(2$^3$S) normalized transmission spectrum in comparison with the double Gaussian fit and TPM simulation.}
% \label{fig:HeI3_fit}
% \end{figure*}

\begin{table*}
\centering
\renewcommand{\arraystretch}{1.0}
\fontsize{10}{13}\selectfont
\caption{HD 149026b transmission spectroscopy upper limits.}
\label{table:4}
\begin{tabular}{
m{2.0cm}
P{3.5cm}
c
% P{2.0cm}
P{1.5cm}
P{2.0cm}
}

\hline
\hline
% -------------------------
Element       &
Theoretical line in Air &
Flux in the core of the line    &
% Detection                 &
Upper limit (1$\sigma$)       &
Instrument \\
% -------------------------
&
[\r{A}]   &
[\%]        &
% &
[\%]    &
        \\

% -------------------------
% \hline
% CsII        &
% 4603.79   &
%             &
% % N       &
%             &
% HARPS-N            \\
% -------------------------
\hline
\multirow{3}{2.5cm}{\MgI} & 5,167.32 & 16.01 &  0.86 & \\ 
& 5,172.68 & 12.71 & 0.76 & HARPS-N  \\ 
& 5,183.60 & 11.71 & 0.65 & \\
% -------------------------
\hline
\rm Na\,$\scriptstyle\rm I$ D2      &
5,889.95    &
10.03       &
0.45        &
 HARPS-N            \\
% -------------------------
\hline
\rm Na\,$\scriptstyle\rm I$ D1      &
5,895.92    &
12.15       &
0.40        &
 HARPS-N            \\
% -------------------------

\hline
H$\alpha$   &
6,562.81     &
16.70       &
% N         &
0.55        &
 HARPS-N            \\
% -------------------------
\hline
\rm Li\,$\scriptstyle\rm I$    &
6,707.76 &
89.44       &
0.29       &
 HARPS-N        \\
% -------------------------
\hline
\multirow{3}{2.5cm}{\HeI(2$^3$S)} & 10,829.09 & 100 & 0.23 & \\ 
& 10,830.25 & \multirow{2}{*}{96.23} & \multirow{2}{*}{0.34} & GIANO-B  \\ 
& 10,830.34\\
% -------------------------
\hline
\hline

    \end{tabular}
\end{table*}
%-------------------------------------------------------------------

\subsection{Cross-correlation analysis}\label{CCF}

\begin{figure*}
\sidecaption
  \includegraphics[width=12cm]{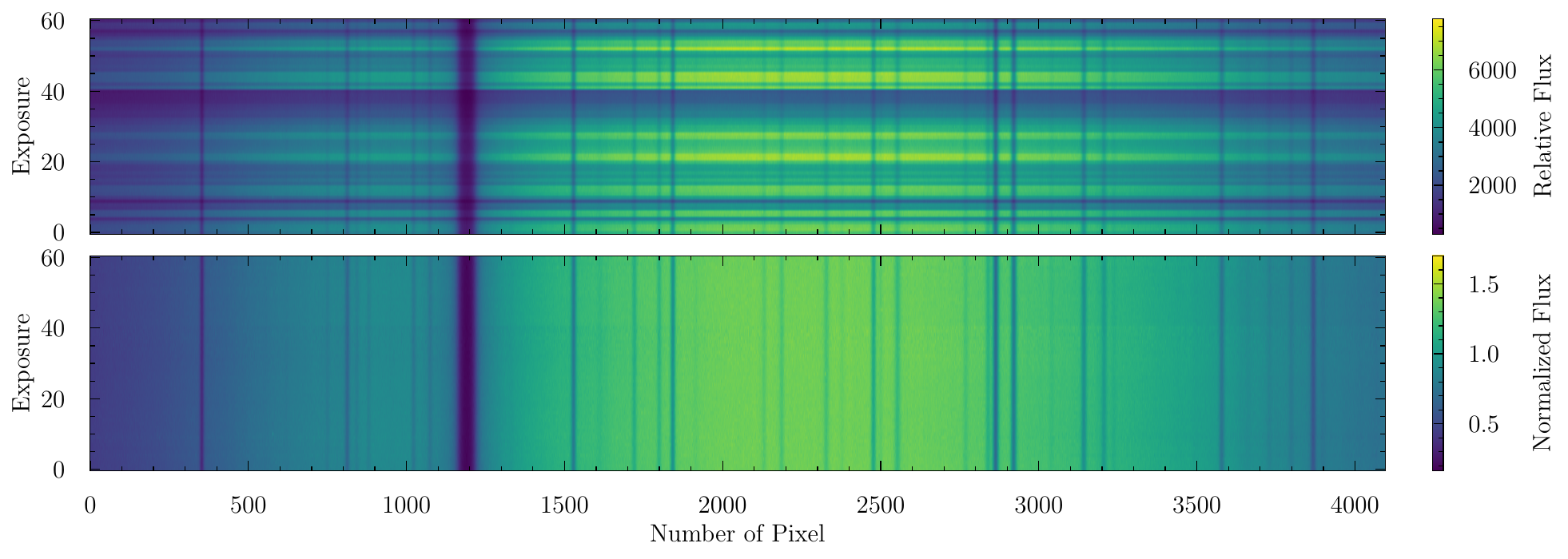}
     \caption{Flux variation for all the exposures for the 59$^{\mathrm{th}}$ HARPS-N spectral order (6,538.7 - 6,612.3~\AA{}). The strong absorption line at pixel $\simeq$ 1,200 is the stellar H$\alpha$ line. Top panel (before the normalisation) shows the 59$^{\rm th}$ spectral order along all the exposures. Bottom panel shows the result after the normalisation.}
\label{fig:ordini_reformat}
\end{figure*}

% \begin{figure*}
% \centering
% \includegraphics[width=1\textwidth]{Figure/lista_exp_ordine_59_normalizzata.pdf}
% \caption{Flux variation for all the exposures for the 59$^{\mathrm{th}}$ HARPS-N spectral order (6,538.7 - 6,612.3~\AA{}). The strong absorption line at pixel $\simeq$ 1,200 is the stellar H$\alpha$ line. Top panel (before the normalisation) shows the 59$^{\rm th}$ spectral order along all the exposures. Bottom panel shows the result after the normalisation.}
% \label{fig:ordini_reformat}
% \end{figure*}

Line-by-line transmission spectroscopy is very effective in detecting absorption lines with large cross-sections in exoplanetary atmospheres. However, it often fails in detecting the numerous, intrinsically faint lines expected from atoms such as, e.g., $\rm Fe$, $\rm Ti$, $\rm V$, or from  molecules. In order to increase SNR, the cross correlation function (CCF) is indeed a more powerful tool \citep[e.g.,][]{Snellen_2010,Brogi_2012,Hoeijmakers_2019,Borsa_2021}. 

In our analysis, we employed the CCF technique to investigate the presence of \TiI, \VI, \CrI, \FeI\, and VO in the visible band, and CH$_4$, CO$_2$, H$_2$O, HCN, NH$_3$, and VO in the nIR. For this purpose, we constructed the templates employing {\fontfamily{pcr}\selectfont petitRADTRANS} \citep{Molliere_2019}, a {\fontfamily{pcr}\selectfont python} package that allows to calculate the planetary transmission and emission spectra. We constructed models by using the parameters of Table \ref{table:2}, with an isothermal atmospheric profile with T = 2000 K, a metallicity [Fe/H] = 0.36 \citep{Ishizuka_2021} and a continuum pressure of 10 mbar, which is consistent with the white light transit depth. The models, generated as expected variation in planetary radius with wavelength, were then translated in ($R_{p}/R_\star$)$^2$ and convolved with the HARPS-N resolving power. 
We normalized the templates %to a zero continuum 
and convolved them also to match the planet's rotational velocity, assuming tidal locking.

For the CCF analysis, we started performing similar steps as done for the line-by-line analysis on the s1d spectra, but using the e2ds spectra \citep[e.g.,][]{Stangret_2020}. This allows us to perform a better normalization when working on the whole spectral range of the spectrograph.
We employed {\fontfamily{pcr}\selectfont Molecfit} to correct the 1D s1d HARPS-N spectra and applied the retrieved telluric profile to correct the e2ds spectral orders \citep[e.g.,][]{Hoeijmakers_2020}. To enhance the quality of the CCF analysis, we excluded the first five orders and the last one from all exposures due to their low SNR, focusing on the wavelength range 4,000-6,840~\AA{}.
The e2ds spectral orders were then reformatted into matrices, each of them contain all the exposures of the same order (Fig. \ref{fig:ordini_reformat}, top panel). As we are working with ground-based instruments, we must account for the flux variation over time. Therefore, we normalized each exposure by dividing it by the mean value across the pixels (Fig. \ref{fig:ordini_reformat}, lower panel). 

\begin{figure*}
\sidecaption
\raisebox{3.5cm}{
\begin{minipage}{12cm}
  \includegraphics[width=\textwidth]{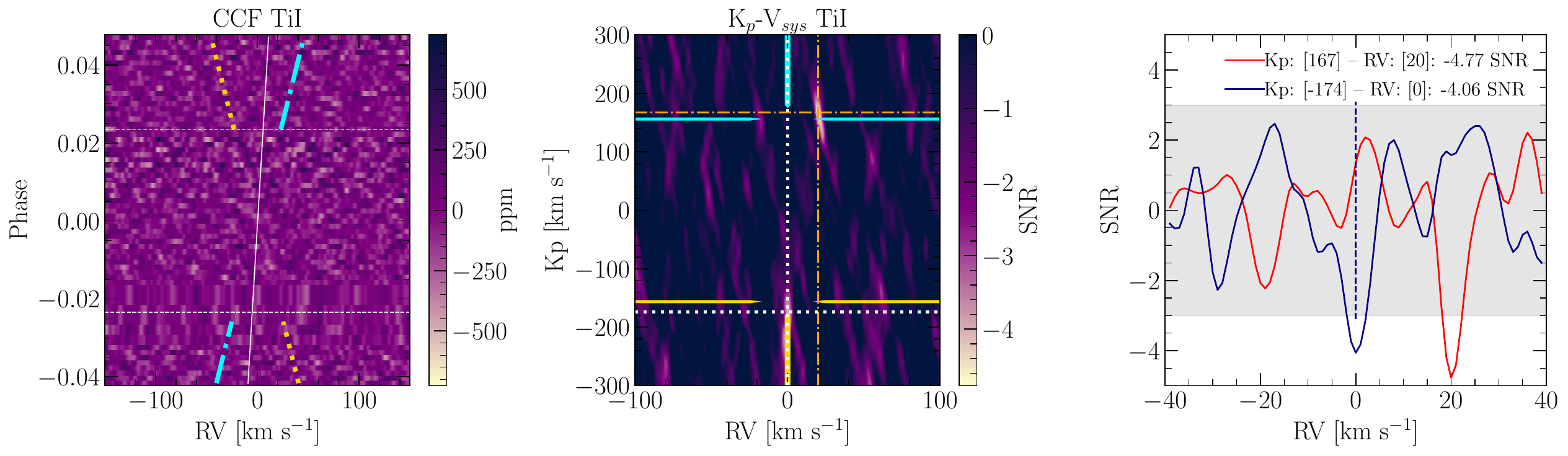}\\
  \includegraphics[width=\textwidth]{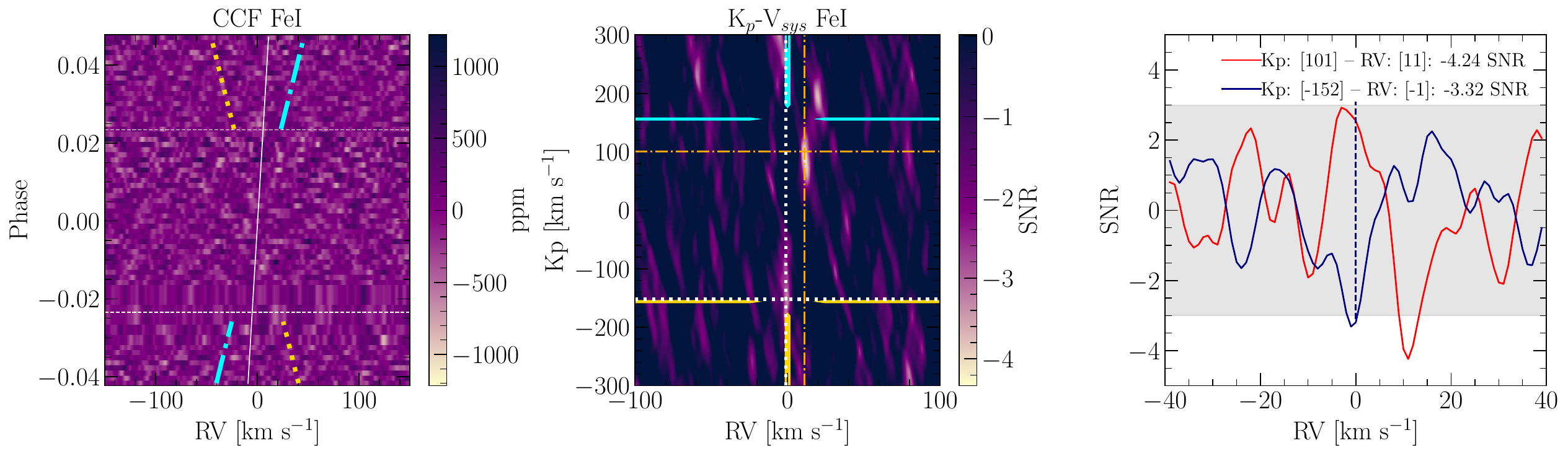}
\end{minipage}
}
     \caption{CCF (left) and $K_p - V_{\rm sys}$ (center) maps for \TiI \, and \FeI. The horizontal white lines in CCF maps correspond to the T$_1$ and T$_4$ contact points, while the cyan dash dotted and gold dotted lines mark the expected planetary $K_p$ and -$K_p$ of the injected signals, respectively. The white slanted line corresponds to the Doppler shadow curve. Cyan and gold arrows in $K_p - V_{\rm sys}$ maps aim at $K_p$ and -$K_p$, respectively. Orange dash dotted and white dotted pointers aim respectively at the minimum value of the maps between RV [-40, +40]~\kms,\, $K_p$ [100,200]~\kms and between RV [-40, +40]~\kms,\, $K_p$ [-200,-100]~\kms.  The red and navy lines in the last column are the 1D projection of the $K_p - V_{\rm sys}$ maps evaluated at the minima found around the theoretical and injected $K_p$.}\label{fig:CCF_Kp}
\end{figure*}

\begin{figure*}
\sidecaption
% \raisebox{3.5cm}{
\begin{minipage}{12cm}
  \includegraphics[width=\textwidth]{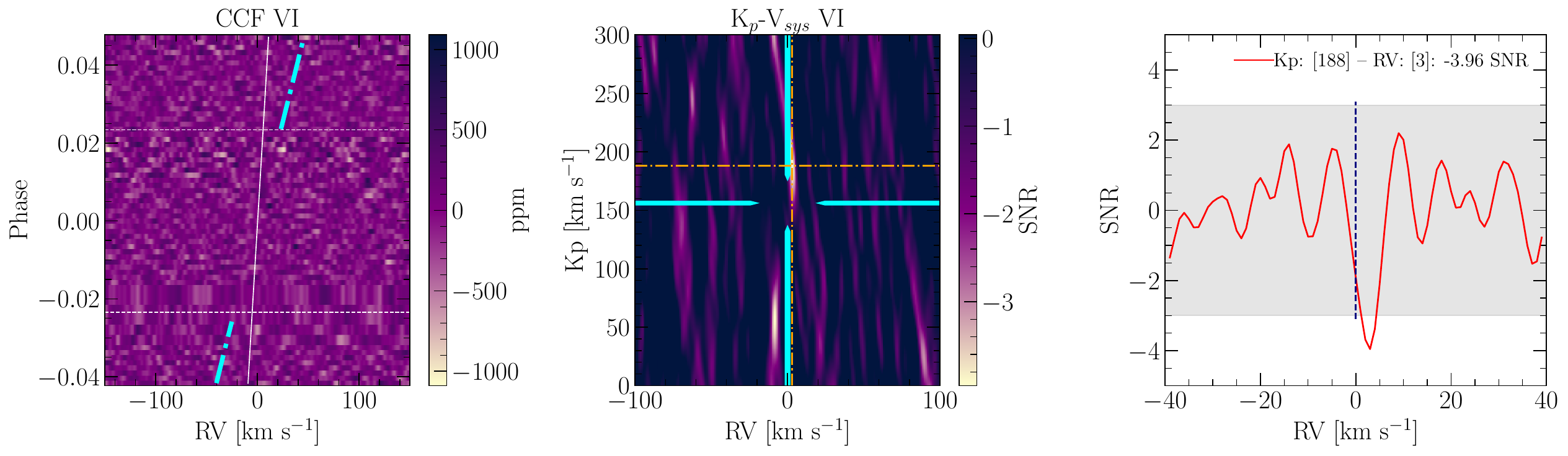}\\
  \includegraphics[width=\textwidth]{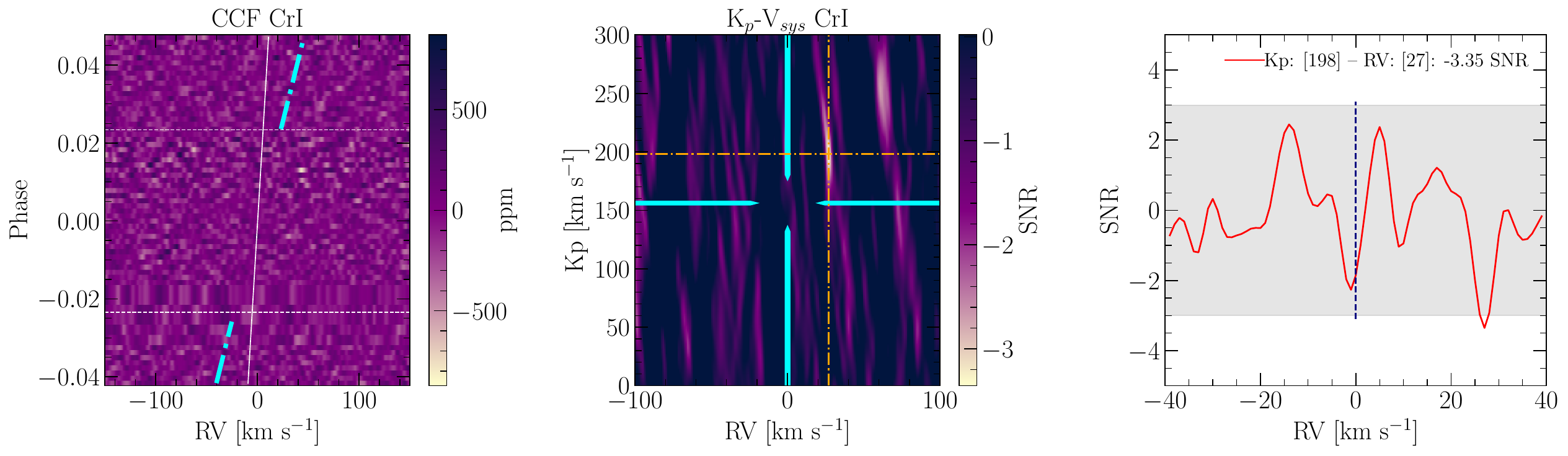}
\end{minipage}
% }
\caption{CCF results for \VI\, and \CrI\, as in Fig. \ref{fig:CCF_Kp}}
\label{fig:CCF_Kp_2} 
\end{figure*}

We then started our cross-correlation function calculation,

\begin{equation}
    CCF(v,t) = \frac{\sum_{k}{x_k(t) \cdot T_k(v)}}{\sum_k T_k(v)},
\end{equation}

where $v$ is the velocity at which the template is shifted and $k$ refers to the pixels in our one-dimensional spectrum $x$ that contains all the stacked normalised and telluric corrected spectral orders, while $T$ represents the template. 
The CCF is done in the range -200~\kms < v < 200~\kms, in steps of 1~\kms.
The resulting CCF was shifted to the stellar reference frame, interpolated within the RV range -150~\kms < v < 150~\kms in steps of 1~\kms and normalized by the CCF $M_{\rm out}$. For a more accurate normalization, we further divided this CCF by the median values across the exposures.
To eliminate low-frequency fluctuations and smoothing the CCF, we conducted a Fast Fourier Transform cutting any velocity beyond 100~\kms for each exposure. 
The resulting final CCFs for the inspected elements are given in Fig.~\ref{fig:CCF_Kp} and Fig.~\ref{fig:CCF_Kp_2}. The blurred region at phase $\simeq\,-0.02$ is caused by the lack of data, since we discarded them to their low SNR.

To detect any possible planetary atmospheric signal, we generated the $K_p - V_{\rm sys}$ maps by shifting all the CCFs in different planetary reference systems. We considered circular orbits with various $K_p$ velocities in the range 0-300~\kms, with a step of 1~\kms, and averaged all the in-transit residual CCFs. 
$K_p - V_{\rm sys}$ maps are then divided by the standard deviation of the overall map, after excluding a region of $\pm 100$~\kms for both RV and $K_p$ values centered at the expected $K_p$ value (156.19~\kms) and 0~\kms\, RV. This normalization yields $K_p - V_{\rm sys}$ maps in terms of SNR as shown in Fig.~\ref{fig:CCF_Kp} and Fig.~\ref{fig:CCF_Kp_2}. The blurred regions in the CCFs at the beginning of the transit are due to the lack of data for these orbital phases (see Fig.~\ref{fig:harpsn_gianob_snr} and Sect. \ref{Observations}).

From the analysis of our data, we do not find evidence for the presence of any of the searched species (SNR $> 4$) in the atmosphere of the planet in proximity of the expected planetary signal. The non-detection of \TiI, in particular, is in contrast with the previous detection of this element by \citet{Ishizuka_2021}.
In order to verify if with our data we are sensitive to the expected signals, we performed the injection of our planetary atmospheric models into the data prior to any analysis, using a planetary orbital velocity opposite to the theoretical one \citep[e.g.,][]{Pelletier_2021}.
The e2ds spectra were then analysed following the same prescription described at the beginning of Sect. \ref{CCF}. CCF and $K_p - V_{\rm sys}$ maps containing the injected models are shown in Fig.~\ref{fig:CCF_Kp}.
In this way, we retrieved the injected signal of \TiI \, and \FeI \, at the expected $K_p$ and RV value at $\sim 4.1 \, \sigma$ ($K_p = -174$~\kms, RV = 0~\kms) and $\sim 3.3 \, \sigma$ ($K_p = -152$~\kms, RV = $-1$~\kms) respectively. For \VI\, and \CrI\, the data are not sensitive to our planetary model to give us a clear detection.

Since the results of CCF analysis may strongly depend on the line lists used to create the templates \citep[e.g.,][]{Gandhi20}, we conducted a check on our \TiI \, non-detection by creating templates with different line lists.
We used the linelists from Kurucz \citep{Kurucz_1993}, VALD \citep{Piskunov_1995}, and NIST downloaded from the DACE\footnote{dace.unige.ch/opacityDatabase/} database and converted in {\fontfamily{pcr}\selectfont petitRADTRANS} format to create new models, and the isothermal templates at 2000 K, 2500 K, and 3000 K  from \cite{Kitzmann_2023}. We then re-performed the CCFs by using all these templates, obtaining very similar results and non-detections each time.

\begin{figure*}
\centering
\includegraphics[width=1\textwidth]{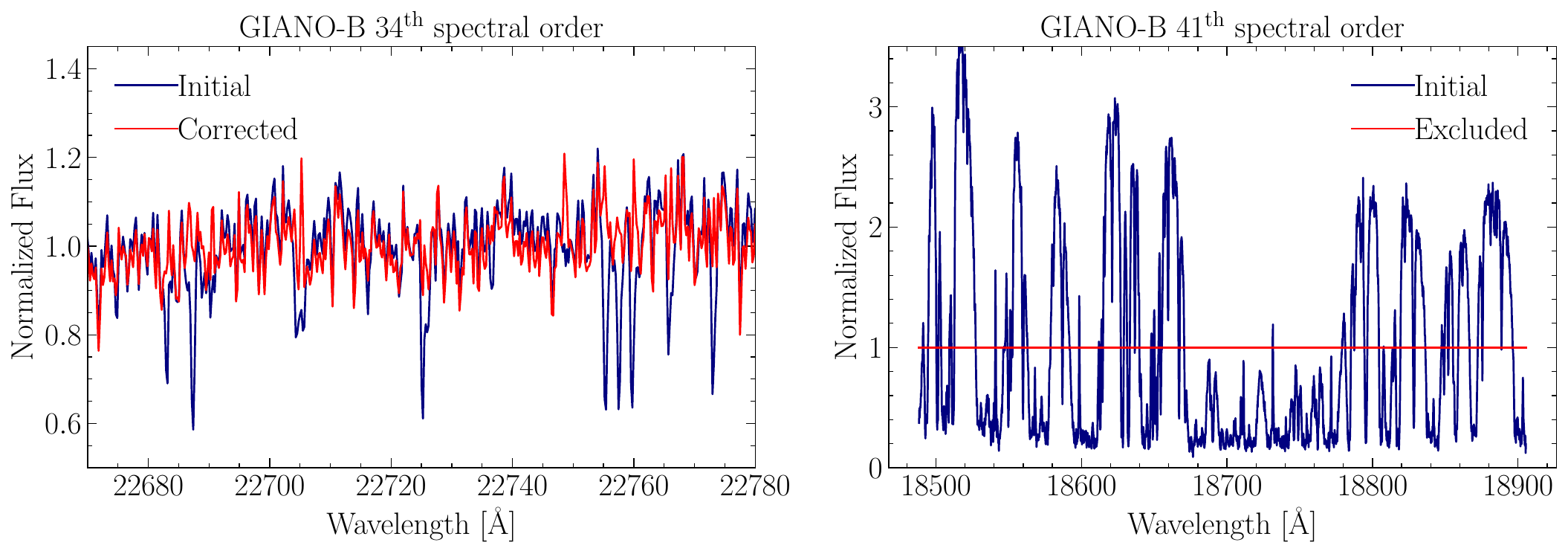}
\caption{Comparison between two different GIANO-B ms1d spectral orders, before (navy) and after (red) {\fontfamily{pcr}\selectfont Molecfit} correction. Left panel refers to the 34$^{\rm th}$ spectral order, where the correction is successfull. Right panel refers to the 41$^{\rm th}$ spectral order, where correction does not work properly as in this wavelength region telluric lines are saturated. We didn't consider the 41$^{\rm th}$ spectral order and similar ones in our analyses and discarded them.}
\label{fig:ordini}
\end{figure*}

We performed CCFs also in the infrared channel on the ms1d GIANO-B spectral orders, searching for the presence of molecules such as CH$_4$, CO$_2$, H$_2$O, HCN, NH$_3$, and VO. Also in this channel we performed telluric correction using {\fontfamily{pcr}\selectfont Molecfit}, with a dedicated procedure as described in Sect. \ref{Transmission_spectroscopy}. Many orders are not usable due to their low SNR and the intrinsic transmissivity of the Earth's atmosphere.
Therefore, we masked the flux of the following GIANO-B diffraction orders
(81, 80, 79, 69, 68, 67, 66, 57, 56, 55, 54, 53, 52, 51, 43, 42, 41, 40, 39, 38, 37, 32) before the CCF analysis for each exposure. Figure~\ref{fig:ordini} shows an example of two different orders where the {\fontfamily{pcr}\selectfont Molecfit} telluric correction was applied successfully and not, respectively. 

\begin{figure*}
\sidecaption
% \raisebox{3.5cm}{
\begin{minipage}{12cm}
  \includegraphics[width=\textwidth]{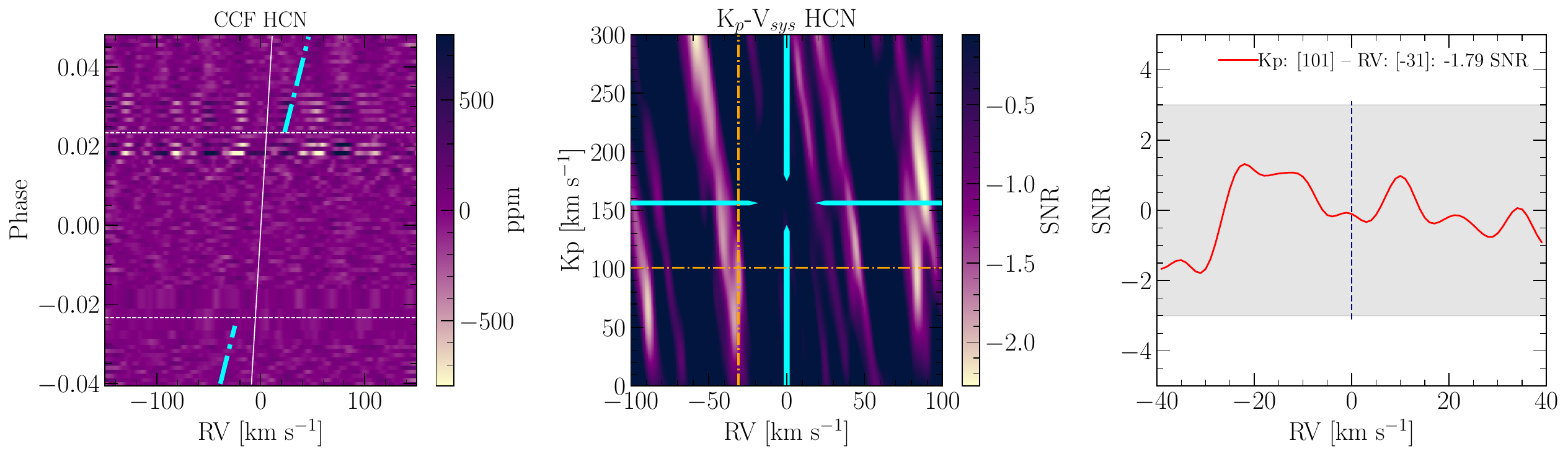}\\
  \includegraphics[width=\textwidth]{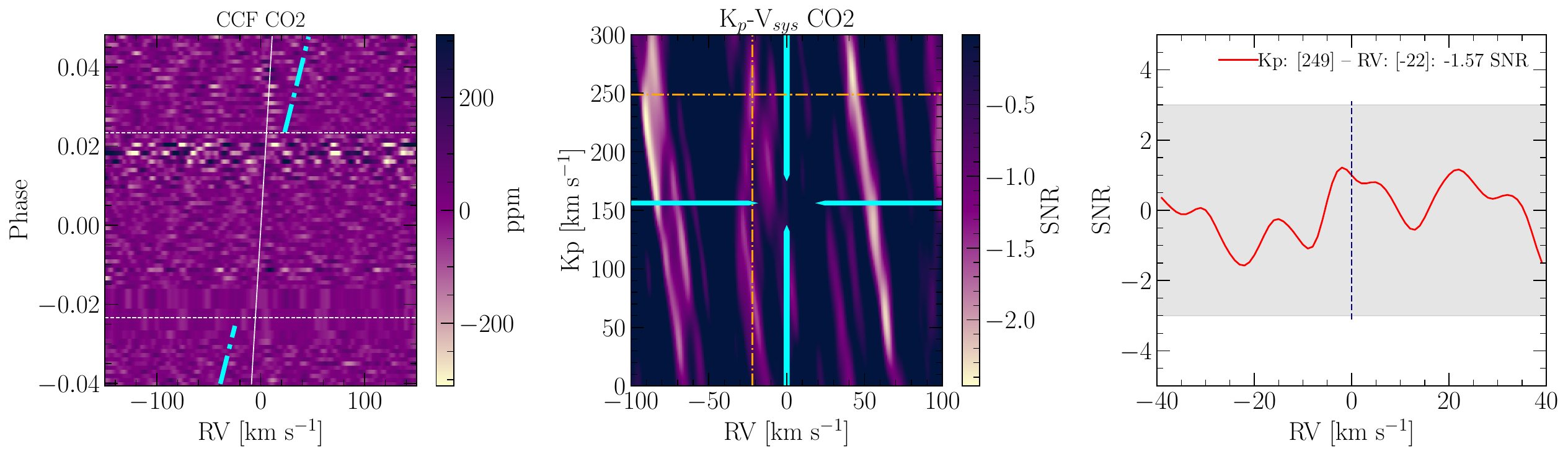}
\end{minipage}
% }
\caption{CCF results for HCN and CO$_{\rm 2}$ as in Fig. \ref{fig:CCF_Kp}. At the end of the transit (phases $\simeq$ 0.02) the noise in the CCF maps is due to the SNR drops for these exposures (see Fig. \ref{fig:harpsn_gianob_snr}).}
\label{fig:CCF_Kp_nIR}
\end{figure*}

The CCF and $K_p - V_{\rm sys}$ maps for the selected models where then retrieved with the same normalisation procedure employed in the VIS band with the HARPS-N e2ds spectral orders. Also in the nIR band, our analysis did not show any detection of exoplanetary absorption signal (Fig.~\ref{fig:CCF_Kp_nIR}).\\

%%%%%%%%%%%%%%%%%%%%%%%%%%%%%%

% \begin{figure*}
% \centering
% \includegraphics[width=0.95\textwidth]{Figure/TiI_final_referee.pdf}\vspace{0.5cm}
% \includegraphics[width=0.95\textwidth]{Figure/FeI_final_referee.pdf}
% \caption{CCF (left) and $K_p - V_{\rm sys}$ (center) maps for \TiI \, and \FeI. The horizontal white lines in CCF maps correspond to the T$_1$ and T$_4$ contact points, while the cyan dash dotted and gold dotted lines mark the expected planetary $K_p$ and -$K_p$ of the injected signals, respectively.
% The white slanted line corresponds to the Doppler shadow curve. Cyan and gold arrows in $K_p - V_{\rm sys}$ maps aim at $K_p$ and -$K_p$, respectively. Orange dash dotted and white dotted pointers aim respectively at the minimum value of the maps between RV [-40, +40]~\kms,\, $K_p$ [100,200]~\kms and between RV [-40, +40]~\kms,\, $K_p$ [-200,-100]~\kms. 
% The red and navy lines in the last column are the 1D projection of the $K_p - V_{\rm sys}$ maps evaluated at the minima found around the theoretical and injected $K_p$.}
% \label{fig:CCF_Kp}
% \end{figure*}
%%%%%%%%%%%%%%%%%%%%%%%%%%%%%%

%%%%%%%%%%%%%%%%%%%%%%%%%%%%%%

% \begin{figure*}
% \centering
% \includegraphics[width=0.95\textwidth]{Figure/VI_final_referee.pdf}\vspace{0.5cm}
% \includegraphics[width=0.95\textwidth]{Figure/CrI_final_referee.pdf}\vspace{0.5cm}
% \caption{CCF results for \VI\, and \CrI\, as in Fig. \ref{fig:CCF_Kp}}
% \label{fig:CCF_Kp_2}
% \end{figure*}
%%%%%%%%%%%%%%%%%%%%%%%%%%%%%%

%%%%%%%%%%%%%%%%%%%%%%%%%%%%%%

% \begin{figure*}
% \centering
% \includegraphics[width=0.95\textwidth]{Figure/HCN_final_referee.pdf}\vspace{0.5cm}
% \includegraphics[width=0.95\textwidth]{Figure/CO2_final_referee.pdf}\vspace{0.5cm}
% \caption{CCF results for HCN and CO$_{\rm 2}$ as in Fig. \ref{fig:CCF_Kp}. At the end of the transit (phases $\simeq$ 0.02) the noise in the CCF maps is due to the SNR drops for these exposures (see Fig. \ref{fig:harpsn_gianob_snr}).}
% \label{fig:CCF_Kp_nIR}
% \end{figure*}
%%%%%%%%%%%%%%%%%%%%%%%%%%%%%%

%%%%%%%%%%%%%%%%%%%%%%%%%%%%%%

\section{Lithium in the stellar spectrum}\label{lithium}

\begin{figure}
\centering
\includegraphics[width=0.5\textwidth]{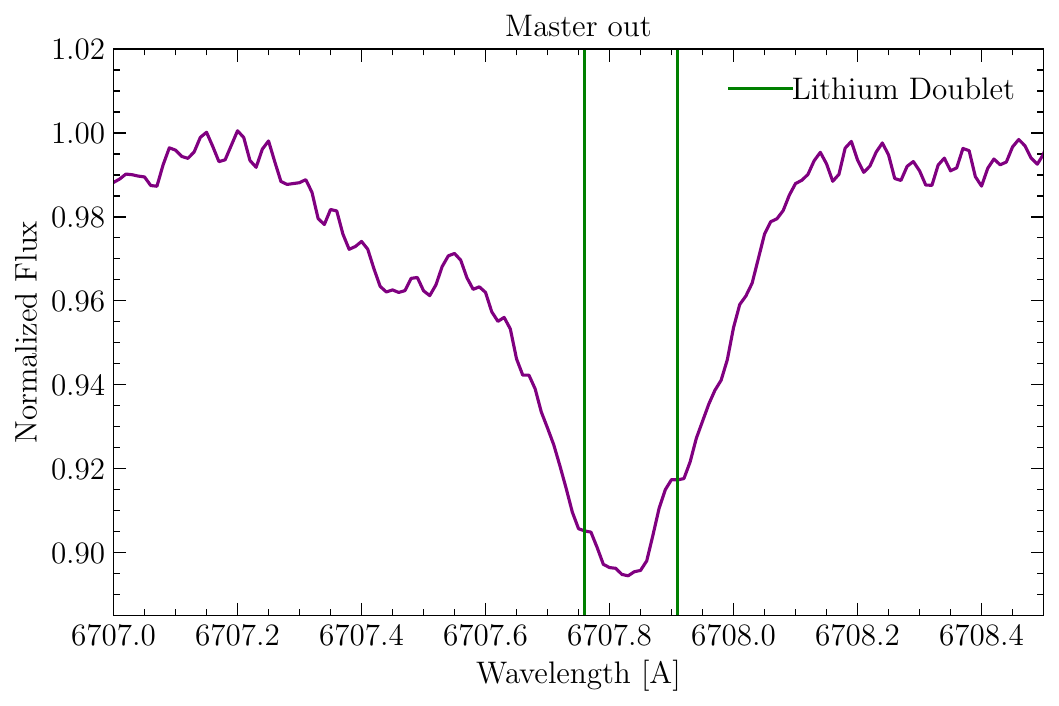}
\caption{Master-out stellar spectrum in proximity of the lithium doublet at 6,707.76 \AA{}.}
\label{fig:litio}
\end{figure}

When investigating planetary atmospheric signals, we also explored the lithium line at 6,707.76 \AA{} (Sect.~\ref{single_line_analyses}). While we did not find it in the planetary atmosphere, we noticed that it is clearly present in the master-out stellar spectrum (Fig.~\ref{fig:litio}), which is not often the case for stars of this age because of its rapid depletion \citep[e.g.,][]{Herbig65}. Different age estimates for this system in the literature are in the range 1.9-2.6 Gyr, all estimated based on evolutionary tracks \citep{Sato_2005,Torres_2008,Carter_2009,Bonomo_2017,Ment_2019}, with just one single exception \citep[1.2 Gyr,][]{Southworth_2010}.
However there is also the possibility of Li enrichment by, e.g., the engulfment or accretion of a close-orbiting substellar companion \citep[e.g.,][]{Soares-Furtado21}.
Interestingly, \citet{Li_2008} already suggested that the observed high metallicity of HD~149026 may be confined to its surface layer as a consequence of pollution by the accretion of a gas giant or a of a population of smaller-mass rocky planets.

Since lithium is often a good age estimator \citep[e.g.,][]{Herbig65,Jeffries23}, 
we investigated the possibility that its presence in the stellar spectrum could be at odds with the estimated age of the parent star.
We estimated the equivalent width of the lithium 6,707.76 \AA{} line to be $36 \pm 2$ m\AA{}, after deblending it with the close \FeI\, line \citep[e.g.,][]{Jeffries23}. We then inserted this value, together with the stellar temperature of Table~\ref{table:2}, in the publicly available EAGLES code \citep{Jeffries23}, which can give an independent stellar age estimate based on these two values.
Unfortunately the results are totally unconstrained, likely because HD~149026 is too hot to use lithium as indicator of stellar age.
One other option to test the engulfment hypothesis would be to look at the \isotope[6]{Li}/\isotope[7]{Li} isotopic ratio \citep[e.g.,][]{Biazzo22,Cuntz00}, but unfortunately our master-out stellar spectrum does not have sufficient SNR.
%La questione del planet engulfment da litio è innanzi tutto a momento una cosa ancora non accettata da tutti (appunto perché spesso i claims fatti sul litio erano in realtà compatibili con lo stadio evolutivo del litio per la specifica target per la quale il claim era stato fatto), ma anche volendo pensare una cosa del genere, l'eventuale presenta di sovrabbondanza di litio legato all'engulfment non è sufficiente perché andrebbero guardati i rapporti isotopici di Li6/Li7 che però hanno bisogno di spettri  di altissimo S/N (anche 500-600). Io ad esempio ci ho provato nel mio lavoro del 2022 ma non sono riuscita ad avere risultati compatibili con scenari di possibile planet engulfment nonostante avessi anche S/N di 100-300. In ogni caso, al di là di questo, la detection del litio trovata da voi non è anomala e compatibile con lo stadio evolutivo in cui si trova la target per una età di qualche miliardo di anni.

%from Biazzo et al. 22: Standard and non-standard stellar evolution models predict in solar-type stars a destruction of 6Li at the base of the convective envelope (Talon & Charbonnel 2005, and references therein), and hence the presence of 6Li in the atmosphere of a planet-hosting star has been justified as indication of an external pollution process, like planetary material accretion or superflares around stars with hot Jupiters (e.g., Cuntz et al. 2000;Israelian et al. 2001; Mott et al. 2017).

\section{Discussion and conclusion} \label{Discussion}

In this work, we analysed VIS and nIR high-resolution spectroscopy observations of one transit of the exoplanet HD~149026b. After analyzing the Rossiter-McLaughlin effect, which allowed us to refine the projected spin-orbit angle of the planet, we performed transmission spectroscopy. 
We searched for atomic and molecular species in the atmosphere of the planet by using both line-by-line and CCF techniques. We could not detect any of the searched species, finding only upper limits (Table~\ref{table:4}). Indeed, the low flux in the core of the stellar lines, like it is common for G-type stars, does not help the detection of small atmospheric signals.

It is worth noticing that the $K_p - V_{\rm sys}$ maps for \TiI, \FeI, and to a lesser extent \CrI, may hint to a statistically significant (SNR=4.7, 4.2 and 3.3 respectively) signal at $\simeq +(10-27)$~\kms relative to the expected position in radial velocity, consistent with the planetary $K_p$ value (Fig.~\ref{fig:CCF_Kp} and \ref{fig:CCF_Kp_2}). 
In the literature there has been some debate regarding a possible eccentricity of the HD~149026b orbit, which could indeed bring such a velocity shift for a signal from the planetary atmosphere. \citet{Wang_2011} found an eccentricity $e\simeq 0.19 \pm 0.07$. \citet{Harrington_2007} detected an earlier occurrence of the secondary eclipse using Spitzer Space Telescope data, approximately at -3 minutes the expected timing. Similarly, analyzing data obtained three years after, \citet{Knutson_2009} predicted a secondary eclipse occurring approximately -21 minutes earlier than expected, possibly suggesting an eccentric orbit ($e \cos \omega \simeq -0.0079$ where $\omega$ is the argument of pericenter). However, the most recent analyses of both RVs and light curves of HD~149026b seem to be consistent with no eccentricity \citep{Bonomo_2017,Bean_2023}. 
To further check if even a small eccentricity could cause this shift in any atmospheric signal, we verified that the values of 
% $e = 0.051 \pm 0.019$ \, and $\omega = 109 \pm 21$
$e$ and $\omega$ parameters from \citet{Ment_2019}, when varying them within errorbars, can not be the source of a shift this large. 
By calculating the radial velocity (RV) curves with these parameters, we find curves that have a maximum RV of +6 \kms at mid-transit (phase = 0). When varying the parameters of $e$ and $\omega$ within 3 $\sigma$, the most extreme curve shows an RV of +15.5 \kms at phase = 0, which is still different from the observed signal at +20 \kms. We thus tend to exclude orbital eccentricity as the possible cause of these signals.
An alternative, fascinating origin for this highly redshifted ($\simeq +20$~\kms) signal is from material falling on the star. \citet{Li_2008} already suggested that this star may be highly metallic as a consequence of pollution by the accretion of planets. We investigated this hypothesis by studying the stellar lithium, but we found inconclusive results. Since we have only one transit observation and cannot confirm it independently, we thus tend to attribute this signal to a spurious fluctuation. We however suggest that this should be further investigated with multiple high-resolution transit observations.

\citet{Ishizuka_2021} reported a $\simeq 4.4\,\sigma$ detection of \TiI, together a tentative, $\simeq 3\,\sigma$ detection of \FeI. They assumed a circular orbit. Contrary to \citet{Ishizuka_2021}, our analysis did not detect either \TiI~nor \FeI~in the data in the planetary restframe. By performing injection-retrieval tests, we showed that our dataset is sensitive to the model of \TiI \, injected (SNR=4.1), and partially to the \FeI \, one (SNR=3.3). Our analysis of the HARPS-N dataset thus excludes the presence of \TiI \, in the atmosphere of the planet, in contrast with \citet{Ishizuka_2021}.

The actual absence of \TiI~in the planetary atmosphere could be in line with the possible presence of the so-called “titanium cold trap”. The cold trap occurs when, as in the case of HD~149026b, the equilibrium temperature is below the threshold for observing titanium in the upper atmosphere. According to \citet{Hoeijmakers_2022}, in planets with temperatures below $\simeq$ 2200 K, \TiI~condenses on the night side. This condensation could result in its absence throughout the planetary atmosphere, rendering it undetectable in the upper atmospheric layers where transmission spectroscopy is most effective to probe. The cold-trap process can cause titanium to remain trapped on the planet's night side due to inefficient advection. Alternatively, if titanium is circulated back into the day side, it may reside in high-pressure, low altitude layers where it is hardly detectable. In conclusion, the low equilibrium temperature and absence of 
\TiI ~in the planet's atmosphere support the hypothesis of a titanium cold trap. Alternatively, the planet could possess a very dense atmosphere with extensive cloud coverage, making the atmosphere optically thick altogether at optical and near-infrared wavelengths.

\begin{acknowledgements}
  We thank K. Biazzo and V. D'Orazi for useful suggestions on the presence of lithium in the stellar spectrum. "FB acknowledges support from Bando Ricerca Fondamentale INAF 2023. This work has made use of the VALD database, operated at Uppsala University, the Institute of Astronomy RAS in Moscow, and the University of Vienna. We thank G. Fedrigo and M. Pozzarelli for advice on choosing colors for color-blind in CCF and K$_p - $V$_{\rm sys}$ figures.
\end{acknowledgements}

%-------------------------------------------------------------------
\bibliographystyle{aa} % style aa.bst
\bibliography{bibliography.bib}

\begin{appendix}

\section{Rossiter-McLaughlin effect fitting}\label{appendice}
% \begin{appendix}
\begin{table}[ht!]
\centering
\caption{set of priors used for the RM analysis.}\label{table:3}
% \renewcommand{\arraystretch}{1.5}
% \fontsize{13}{12}\selectfont
% \resizebox{13cm}{3cm}{
\begin{tabular}{lcc}
% m{2.5cm}
% P{1.5cm}
% P{1.5cm}
% }
\hline
\hline
% -------------------------
Uniform Parameters       &
min   &
max   \\ 
% -------------------------
\hline
Systemic velocity$^{(a)}$ $\gamma \,\, $[\kms]      &
-18.5                  &
-17.5                  \\
    % -------------------------
% $\log{\sigma_{w}}$ &
% -10 &
% -1 \\

% -------------------------
\hline
\hline
Gaussian Parameters  &
$\mu$ &
$\sigma$ \\
% -------------------------
\hline
Radial velocity$^{(b)}$ $K_{\star} \,\, $[\kms] &
0.0379  &
0.005   \\

Inclination$^{(b)}$ $i \,\, $[deg]  &
84.50   &
0.6 \\

Stellar rotational velocity$^{(b)}$ $v \sin{i} \,\, $[\kms] &
6.00    &
0.50    \\

$R_p/R_{\star}$ $^{(c)}$ &
0.050565  &
0.0050565  \\

Projected obliquity$^{(d)}$ $\lambda \,\, $[deg] &
12  &
7   \\

$\delta T_0 \,\, [Phase]$ $^{(a)}$ &
0 &
0.1 \\
\hline
\hline

    \end{tabular}
% }
\footnotesize
\begin{flushleft}
(a) This work;
(b) \citet{Bonomo_2017};
(c) \citet{Spinelli_2023};
(d) \citet{Albrecht_2012}.\\
The parameters are distributed with a uniform and a normal distribution. The minimum and maximum values correspond to the boundaries of the uniform distribution, while $\mu$ and $\sigma$ correspond to the center and the standard deviation of the normal distribution.
\end{flushleft}

\end{table}
The RVs were extracted by the HARPS-N DRS using a G2 mask with a CCF width of 20 \kms. We used only the RVs extracted from spectra with SNR $>$ 40 (identified as red dots in Fig. \ref{fig:harpsn_gianob_snr}).
{\fontfamily{pcr}\selectfont CaRM} employs an MCMC algorithm through  {\fontfamily{pcr}\selectfont emcee}, a {\fontfamily{pcr}\selectfont python} module developed by \citet{Foreman_2013}. Table \ref{table:3} provides all the priors used in our analysis. The guess parameter for the systemic radial velocity used, $\gamma = -18.035 \, \rm{km \, s^{-1}}$, was taken from an approximate RVs data estimation. The other guess parameters we used are:  the ratio between the semi-major axis and the stellar radius $a/R_\star = 5.98$ (\citet{Spinelli_2023}), the  width of non-rotating star based on the HARPS-N resolving power (R $\simeq 115,000$) $\beta_{0} = 2.6 \, $ km s$^{-1}$, the width of the best Gaussian fit to the stellar CCFs $\sigma_0 = 4.37 \, $ km s$^{-1}$ determined from a mean out-of-transit (master-out) CCF\footnote{The Gaussian fit was obtained using {\fontfamily{pcr}\selectfont CPNest} \citep{Del_Pozzo_2022}\footnote{\url{https://github.com/johnveitch/cpnest}}, a {\fontfamily{pcr}\selectfont python} implementation of the nested sampling algorithm \citep{Skilling_2006}.}, the macro-turbolence amplitude $\zeta_{t} = 4.53 \, $ km s$^{-1}$ calculated from the \citet{Doyle_2014} calibration valid for our stellar parameters, and the logarithmic jitter amplitude $\log \sigma_w = -11.5$.

\begin{figure*}
    \makebox[\textwidth][c]{%
        \includegraphics[width=1\textwidth]{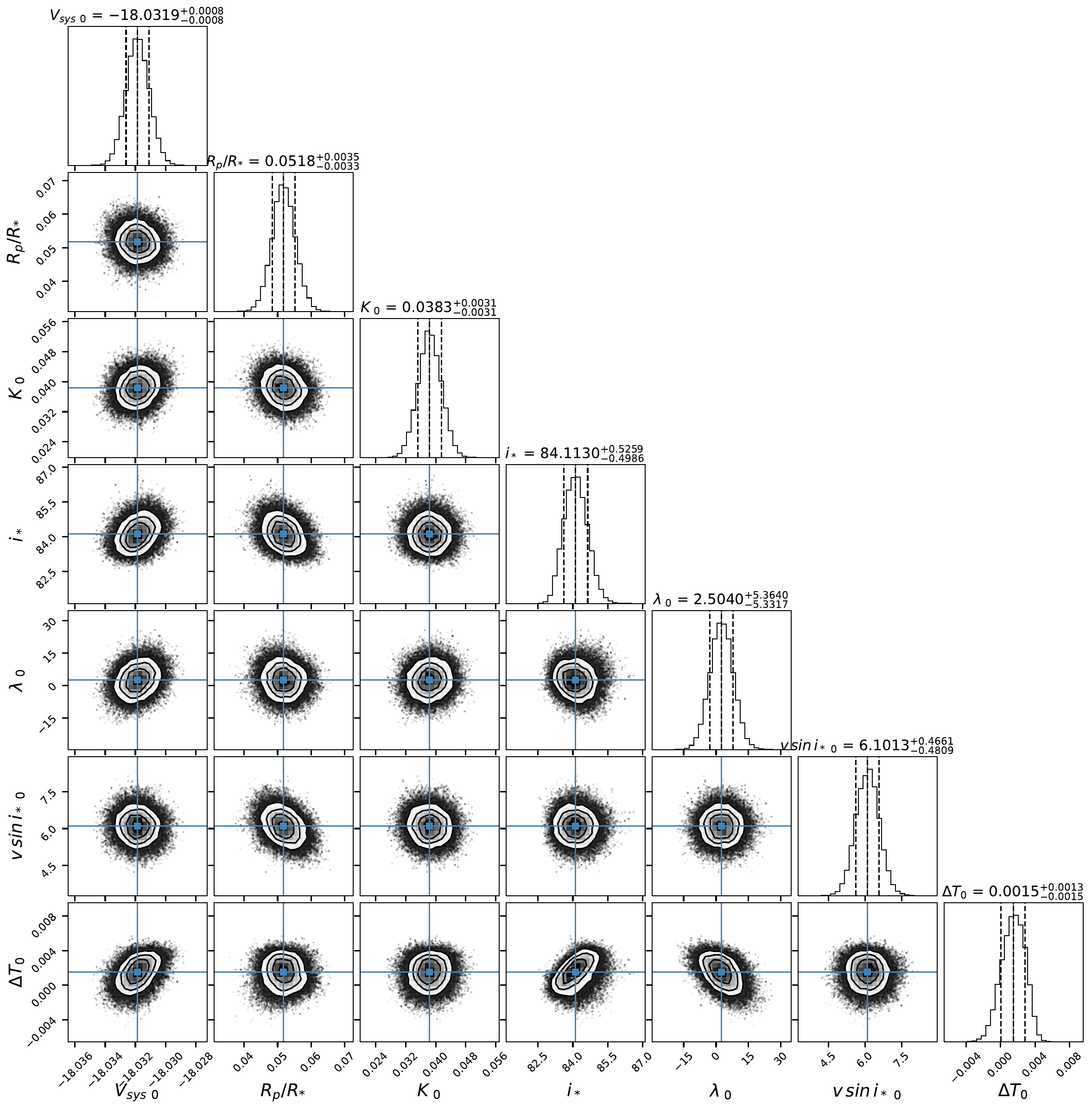} 
    }
    \caption{Posterior distributions of the radial velocities analysis obtained with {\fontfamily{pcr}\selectfont CaRM}.}
\label{fig:posterior}
\end{figure*}

Within {\fontfamily{pcr}\selectfont CaRM}, we specifically employed the model  {\fontfamily{pcr}\selectfont ARoME}, that requires a set of limb-darkening coefficients. {\fontfamily{pcr}\selectfont CaRM} provides them using {\fontfamily{pcr}\selectfont LDTk} \citep{Parviainen_2015}, a {\fontfamily{pcr}\selectfont python} package that automates and models the stellar limb-darkening profiles and coefficients employing spectra from the {\fontfamily{pcr}\selectfont PHOENIX} database \citep{Husser_2013}. In our analysis a quadratic law was employed, yielding the following coefficients: $ldc1 = 0.593$, $ldc2 = 0.123$. Following \citet{Cristo_2022}, we used {\fontfamily{pcr}\selectfont CaRM} with 50 chains, 1,500 steps of burn-in and 3,000 for the production. 
The RVs and their best fit are shown in Fig.~\ref{fig:RM_fit}. The fit matches well the data, except for the initial values which show a trend that is likely caused by the telescope guiding issue detailed in Sect. \ref{Observations}. The results obtained from our analysis are reported in Table~\ref{table:2}, with the posterior distributions shown in Fig.~\ref{fig:posterior}.

\end{appendix}
\end{document}